\documentclass[12pt,aps,superscriptaddress,onecolumn,nofootinbib,preprintnumbers,floatfix]{revtex4}

\usepackage{mathrsfs}
\usepackage{amsfonts,amssymb,amsmath}
\usepackage{graphicx,epsfig}
\usepackage{multirow}
\usepackage{verbatim}

\def\fsu5{$\cal{F}$-$SU(5)$}
\def\bfsu5{$\boldsymbol{\mathcal{F}}$-$\boldsymbol{SU(5)}$}

\def\m1half{$M_{1/2}$}
\def\m3half{$M_{3/2}$}
\def\m32{$M_{32}$}

\def\fb{${\rm fb}^{-1}$~}

\def\mt2{$M_{T2}$}
\def\x2{$\chi^2$}
\def\2b{$M_{T2}b$}
\def\sb{$S/\sqrt{B+1}$~}
\def\bs0{$B_S^0 \rightarrow \mu^+ \mu^-$}

\begin{document}

\title{One-Parameter Model for the Superworld}

\author{Tianjun Li}

\affiliation{State Key Laboratory of Theoretical Physics and Kavli Institute for Theoretical Physics China (KITPC),
Institute of Theoretical Physics, Chinese Academy of Sciences, Beijing 100190, P. R. China}

\affiliation{George P. and Cynthia W. Mitchell Institute for Fundamental Physics and Astronomy, Texas A$\&$M University, College Station, TX 77843, USA}

\author{James A. Maxin}

\affiliation{George P. and Cynthia W. Mitchell Institute for Fundamental Physics and Astronomy, Texas A$\&$M University, College Station, TX 77843, USA}

\author{Dimitri V. Nanopoulos}

\thanks{\normalsize Contribution to the Proceedings of the International School of Subnuclear Physics –- 50th, What we would like LHC to give us, Erice, Sicily, Italy, 23 June –- 2 July 2012, based on a talk given by Dimitri V. Nanopoulos.}

\affiliation{George P. and Cynthia W. Mitchell Institute for Fundamental Physics and Astronomy, Texas A$\&$M University, College Station, TX 77843, USA}

\affiliation{Astroparticle Physics Group, Houston Advanced Research Center (HARC), Mitchell Campus, Woodlands, TX 77381, USA}

\affiliation{Academy of Athens, Division of Natural Sciences, 28 Panepistimiou Avenue, Athens 10679, Greece}

\author{Joel W. Walker}

\affiliation{Department of Physics, Sam Houston State University, Huntsville, TX 77341, USA}


\begin{abstract}

We review the No-Scale \fsu5 model with extra TeV-scale vector-like flippon multiplets and its associated collider phenomenology in the search for supersymmetry at the LHC. The model framework possesses the rather unique capacity to provide a light CP-even Higgs boson mass in the favored 124--126 GeV window while simultaneously retaining a testably light SUSY spectrum that is consistent with emerging low-statistics excesses beyond the Standard Model expectation in the ATLAS and CMS multijet data.

\end{abstract}


\preprint{ACT-14-12, MIFPA-12-38}

\maketitle


\section{No-Scale $\cal{F}$-$SU(5)$}

\subsection{Phenomenological Overview}

We have demonstrated~\cite{Li:2010ws,Li:2010mi} the unique phenomenological
consistency and profound predictive capacity of a model dubbed
No-Scale ${\cal F}$-$SU(5)$, resting essentially and in equal measure on the tripodal
foundations of the ${\cal F}$-lipped $SU(5)$ Grand Unified Theory
(GUT)~\cite{Barr:1981qv,Derendinger:1983aj,Antoniadis:1987dx}, two pairs of
hypothetical TeV scale vector-like supersymmetric multiplets with origins in
${\cal F}$-theory model building~\cite{Jiang:2006hf,Jiang:2009zza,Jiang:2009za,Li:2010dp,Li:2010rz},
and the dynamically established boundary conditions of No-Scale Supergravity
(SUGRA)~\cite{Cremmer:1983bf,Ellis:1983sf, Ellis:1983ei, Ellis:1984bm, Lahanas:1986uc}.
It appears that the No-Scale scenario, and most stringently the vanishing of the Higgs
bilinear soft term $B_\mu$, comes into its own only when applied at an elevated
scale, approaching the Planck mass.  $M_{\cal F}$, the point of the ultimate second stage
$SU(5)\times U(1)_{\rm X}$ unification, emerges in turn as a suitable candidate scale
only when substantially decoupled from the penultimate GUT scale unification
of $SU(3)_C\times SU(2)_L$ at $M_{32} \simeq 10^{16}$~GeV via the modification to
the renormalization group equations (RGEs) from the extra vector-like multiplets.

We have systematically established the hyper-surface within
the $\tan \beta$, top quark mass $m_{t}$, gaugino mass
$M_{1/2}$, and vector-like particle mass $M_{V}$ parameter
volume which is compatible with the application of the simplest
No-Scale SUGRA boundary conditions~\cite{Cremmer:1983bf,Ellis:1983sf, Ellis:1983ei, Ellis:1984bm, Lahanas:1986uc}.
We have demonstrated that simultaneous adherence to all current experimental
constraints, most importantly contributions to the muon anomalous
magnetic moment $(g-2)_\mu$~\cite{Bennett:2004pv}, the branching ratio limit on
$(b \rightarrow s\gamma)$~\cite{Barberio:2007cr, Misiak:2006zs},
and the 7-year WMAP relic density measurement~\cite{Komatsu:2010fb},
dramatically reduces the allowed solutions to a highly non-trivial
``golden strip'', tightly confining $\tan \beta$, $m_{t}$, $M_{1/2}$, and $M_{V}$, effectively
eliminating all extraneously tunable model parameters, where the consonance of the
theoretically viable $m_{t}$ range with the experimentally established value~\cite{:1900yx}
may be interpreted an independently correlated ``postdiction''.  Finally, taking a fixed $Z$-boson mass,
we have dynamically determined the universal gaugino mass $M_{1/2}$ and fixed $\tan \beta$ via the ``Super No-Scale''
mechanism~\cite{Li:2010uu}, that being the secondary minimization, at a local {\it minimum minimorum},
of the minimum $V_{\rm min}$ of the Higgs potential for the electroweak symmetry breaking (EWSB) vacuum.

This model is moreover quite
interesting from a phenomenological point of view~\cite{Jiang:2009zza,Jiang:2009za}. The predicted
vector-like particles can be observed at the Large Hadron Collider (LHC), though possibly
not during the initial run.  The partial lifetime for proton decay
in the leading ${(e|\mu)}^{+} \pi^0 $ channels falls around
$5 \times 10^{34}$ years~\cite{Li:2010dp,Li:2010rz}, testable at the future
Hyper-Kamiokande~\cite{Nakamura:2003hk} and
Deep Underground Science and Engineering Laboratory (DUSEL)~\cite{Raby:2008pd}
experiments~\cite{Li:2009fq, Li:2010dp, Li:2010rz}.
The lightest CP-even Higgs boson mass can be increased~\cite{Huo:2011zt},
hybrid inflation can be naturally realized, and the
correct cosmic primordial density fluctuations can be
generated~\cite{Kyae:2005nv}.

\subsection{The $\cal{F}$-lipped SU(5) GUT}

Gauge coupling unification strongly suggests the existence of a GUT.
In minimal supersymmetric $SU(5)$ models
there are problems with doublet-triplet splitting and dimension
five proton decay by colored Higgsino exchange~\cite{Antoniadis:1987dx}. These difficulties
can be elegantly overcome in Flipped $SU(5)$ GUT
models~\cite{Barr:1981qv, Derendinger:1983aj, Antoniadis:1987dx}
via the missing partner mechanism~\cite{Antoniadis:1987dx}.

Written in full, the gauge group of Flipped $SU(5)$ is
$SU(5)\times U(1)_{X}$, which can be embedded into $SO(10)$.
The generator $U(1)_{Y'}$ is defined for fundamental five-plets as
$-1/3$ for the triplet members, and $+1/2$ for the doublet.
The hypercharge is given by $Q_{Y}=( Q_{X}-Q_{Y'})/5$.
There are three families of Standard Model (SM) fermions,
whose quantum numbers under the $SU(5)\times U(1)_{X}$ gauge group are
\begin{eqnarray}
F_i={\mathbf{(10, 1)}},\quad {\bar f}_i={\mathbf{(\bar 5, -3)}},\quad
{\bar l}_i={\mathbf{(1, 5)}},
\label{smfermions}
\end{eqnarray}
where $i=1, 2, 3$. 
To break the GUT and electroweak gauge symmetries, we 
introduce two pairs of Higgs fields:
a pair of ten-plet Higgs for breaking the GUT symmetry, and a pair
of five-plet Higgs for electroweak symmetry breaking. 
\begin{eqnarray}
& H={\mathbf{(10, 1)}}\quad;\quad~{\overline{H}}={\mathbf{({\overline{10}}, -1)}} &\\
& h={\mathbf{(5, -2)}}\quad;\quad~{\overline h}={\mathbf{({\bar {5}}, 2)}} &
\label{Higgse1}
\end{eqnarray}

A most notable intrinsic feature of the Flipped $SU(5)$ GUT is the presence of dual unification scales, with the
ultimate merger of $SU(5) \times U(1)_{\rm X}$ occurring subsequent in energy to the penultimate $SU(3)_{\rm C}$ and $SU(2)_{\rm L}$
mixing at $M_{32}$.  In the more traditional Flipped $SU(5)$ formulations, the scale $M_{\cal F}$ has been only
slightly elevated from $M_{32}$, larger by a factor of perhaps only two or three~\cite{Ellis:2002vk}.  Our interest
however, is in scenarios where the ratio $M_{\cal F}/M_{32}$ is considerably larger, on the order of $10$ to $100$.

Key motivations for this picture include the desire to address the monopole problem via hybrid inflation,
and the opportunity for realizing true string scale gauge coupling unification in
the free fermionic model building context~\cite{Jiang:2006hf, Lopez:1992kg},
or the decoupling scenario in F-theory models~\cite{Jiang:2009zza,Jiang:2009za}.
We have previously also considered the favorable effect of such considerations on the decay rate of the proton~\cite{Li:2010dp,Li:2010rz}.
Our greatest present interest however, is the effortless manner in which the elevation of the $SU(5) \times U(1)_{\rm X}$
scale salvages the dynamically established boundary conditions of No-Scale Supergravity.  Being highly predictive,
these conditions are thus also intrinsically highly constrained, and notoriously difficult to realize generically.

\subsection{$\cal{F}$-theory Vector-Like Multiplets}

We have introduced additional vector-like particle multiplets derived
within the $\cal{F}$-theory~\cite{Jiang:2006hf} model building context
to address the ``little hierarchy'' problem, altering the $\beta$-coefficients
of the renormalization group to dynamically elevate the secondary 
$SU(5)\times U(1)_{\rm X}$ unification at $M_{\cal F}$ to near the Planck
scale, while leaving the $SU(3)_C\times SU(2)_L$ unification at $M_{32}$
close to the traditional GUT scale.  In other words,
one obtains true string-scale gauge coupling unification in 
free fermionic string models~\cite{Jiang:2006hf,Lopez:1992kg} or
the decoupling scenario in F-theory models~\cite{Jiang:2009zza,Jiang:2009za}.
To avoid a Landau pole for the strong
coupling constant, we are restricted around the TeV scale
to one of the following two multiplet sets~\cite{Jiang:2006hf}.
\begin{eqnarray}
\hspace{-.3in}
& Z1:~ \left( {XF}_{\mathbf{(10,1)}} \equiv (XQ,XD^c,XN^c),~{\overline{XF}}_{\mathbf{({\overline{10}},-1)}} \right)& \nonumber \\
&  Z2:~\left( {XF}, ~{\overline{XF}},
{Xl}_{\mathbf{(1, -5)}},~{\overline{Xl}}_{\mathbf{(1, 5)}}\equiv XE^c \right) &\label{z1z2}
\end{eqnarray}
In the prior, $XQ$, $XD^c$, $XE^c$, $XN^c$ have the same quantum numbers as the
quark doublet, the right-handed down-type quark, charged lepton, and
neutrino, respectively.  We have argued~\cite{Li:2010mi} that the
feasibly near-term detectability of these hypothetical fields in collider experiments,
coupled with the distinctive flipped charge assignments within the multiplet structure,
represents a smoking gun signature for Flipped $SU(5)$, and have thus coined the term
{\it flippons} to collectively describe them.
In this paper, we consider only the $Z2$ set, although discussion for the $Z1$ set,
if supplemented by heavy threshold corrections, can be similar.

We emphasize that the specific representations of vector-like fields which we currently employ have been explicitly
constructed within the local F-theory model building context~\cite{Jiang:2009zza, Jiang:2009za}.  However, the mass of these fields,
and even the fact of their existence, is not mandated by the F-theory, wherein it is also possible to
realize models with only the traditional Flipped (or Standard) $SU(5)$ field content.  We claim only an inherent consistency of their conceptual
origin out of the F-theoretic construction, and take the manifest phenomenological benefits which accompany the elevation of
$M_{\cal F}$ as justification for the greater esteem which we hold for this particular model above other alternatives.

\subsection{No-Scale Supergravity}

The Higgs boson, being a Lorentz scalar,
is not stable in the SM against quadratic quantum mass corrections
which drive it toward the dominant Planck scale, some
seventeen orders of magnitude above the value required for consistent 
EWSB.  Supersymmetry naturally solves this fine tuning problem
by pairing the Higgs with a chiral spin-$1/2$ ``Higgsino'' partner field, and
following suit with a corresponding bosonic (fermionic) superpartner for all
fermionic (bosonic) SM fields, introducing the full set of quantum counter terms.
Localizing the supersymmetry (SUSY) algebra, which includes the generator of
spacetime translations (the momentum operator), induces general coordinate invariance,
producing the supergravity (SUGRA) theories.

Since we do not observe mass degenerate superpartners for the known SM fields,
SUSY must itself be broken around the TeV scale.
In the traditional framework, supersymmetry is broken in the hidden sector, and the effect is 
mediated to the observable sector via gravity or gauge interactions.
In GUTs with minimal gravity mediated supersymmetry breaking, called mSUGRA,
one can fully characterize the supersymmetry breaking
soft terms by four universal parameters
(gaugino mass $M_{1/2}$, scalar mass $M_0$, trilinear coupling $A$, and
the low energy ratio $\tan \beta$ of up- to down-type Higgs VEVs,
plus the sign of the Higgs bilinear mass term $\mu$.
The $\mu$ term and its bilinear
soft term $B_{\mu}$ are determined
by the $Z$-boson mass $M_Z$ and $\tan \beta$ after
the electroweak (EW) symmetry breaking.

No-Scale Supergravity was proposed~\cite{Cremmer:1983bf,Ellis:1983sf, Ellis:1983ei, Ellis:1984bm, Lahanas:1986uc}
to address the cosmological flatness problem,
and defined as the subset of supergravity models
which satisfy the following three constraints~\cite{Cremmer:1983bf}:
(i) The vacuum energy vanishes automatically due to the suitable
K\"ahler potential; (ii) At the minimum of the scalar
potential, there are flat directions which leave the
gravitino mass $M_{3/2}$ undetermined; (iii) The super-trace
quantity ${\rm Str} {\cal M}^2$ is zero at the minimum. Without this,
the large one-loop corrections would force $M_{3/2}$ to be either
zero or of Planck scale.  The defining K\"ahler potential~\cite{Ellis:1984bm}
\begin{eqnarray}
K &=& -3 {\rm ln}( T+\overline{T}-\sum_i \overline{\Phi}_i
\Phi_i)~,~
\label{NS-Kahler}
\end{eqnarray}
automatically satisfies the first two conditions, while
the third is model dependent and can always be satisfied in
principle~\cite{Ferrara:1994kg}.

In Eq.~(\ref{NS-Kahler}), $T$ is a modulus field, while the
$\Phi_i$ are $N_C$ scalar matter fields which parameterize the
coset space $SU(N_C+1, 1)/(SU(N_C+1)\times U(1))$.
The scalar potential is automatically positive semi-definite,
and has a flat direction along the $T$ field.
The non-compact structure of the symmetry implies that the classical vacuum
is not only constant but actually identical to zero.
Moreover, the simplest No-Scale boundary
conditions $M_0=A=B_{\mu}=0$ are dynamically established,
while $M_{1/2}>0$ is allowed, and indeed required for SUSY breaking. A one-parameter model of similar form has been much studied in the past~\cite{Lopez:1993rm,Lopez:1994fz,Lopez:1995hg} (For a review, see \cite{superworld}).
The CP violation problem and the flavor changing neutral current
problems are automatically solved in turn.
All low energy scales are dynamically generated by quantum corrections,
{\it i.e.}~running under the RGEs, to the classically flat potential.

\subsection{The Stringy Super No-Scale Mechanism}

The fiercely reductionist No-Scale picture inherits
an associative weight of motivation from its robustly generic and natural appearance, for example,
in the compactification of the weakly coupled heterotic string theory~\cite{Witten:1985xb}, compactification of
M-theory on $S^1/Z_2$ at the leading order~\cite{Li:1997sk},
and potentially also directly in F-theory models~\cite{Beasley:2008dc,Beasley:2008kw, Donagi:2008ca, Donagi:2008kj}.

In the simplest stringy No-Scale SUGRA, the K\"ahler
modulus $T$, a characteristic of the Calabi-Yau manifold,
is the single relevant modulus field, the dilaton coupling being irrelevant.
The F-term of $T$ generates the gravitino mass $M_{3/2}$, 
which is proportionally equivalent to $M_{1/2}$.
Exploiting the simplest No-Scale boundary condition at $M_{\cal F}$ and 
running from high energy to low energy under the RGEs,
there can be a secondary minimization, or {\it minimum minimorum}, of the minimum of the
Higgs potential $V_{\rm min}$ for the EWSB vacuum.
Since $V_{\rm min}$ depends on $M_{1/2}$, the universal gaugino mass $M_{1/2}$ is consequently 
dynamically determined by the equation $dV_{\rm min}/dM_{1/2}=0$,
aptly referred to as the ``Super No-Scale'' mechanism;
We have argued by the combined action of this mechanism,
the transmutative role of the the RGEs, and the stabilizing counter-balance of
supersymmetry, that No-Scale $\cal{F}$-$SU(5)$ addresses the various aspects of the
gauge hierarchy problem~\cite{Li:2010uu}.

The three parameters $M_0,A,B_{\mu}$ are once again identically zero at the
boundary because of the defining K\"ahler potential, and are thus known at all other scales as well by the RGEs.  The
minimization of the Higgs scalar potential with respect to the neutral elements of both SUSY Higgs doublets gives two
conditions, the first of which fixes the magnitude of $\mu$.  The second condition, which would traditionally be used
to fix $B_{\mu}$, instead here enforces a consistency relationship on the remaining parameters, being that
$B_{\mu}$ is already constrained.

In general, the $B_{\mu} = 0$ condition gives a hypersurface of solutions cut out from a very large parameter space.
If we lock all but one parameter, it will give the final value.  If we take a slice of two dimensional space, as has been 
described, it will give a relation between two parameters for all others fixed.
In a three-dimensional view with $B_{\mu}$ on the vertical axis, this
curve is the ``flat direction'' line along the bottom of the trench of $B_{\mu}=0$ solutions.  In general, we
must vary at least two parameters rather than just one in isolation, in order that their mutual compensation may transport
the solution along this curve.  The most natural first choice is in some sense the pair of prominent unknown inputs 
$M_{1/2}$ and $\tan \beta$, as demonstrated in Ref.~\cite{Li:2010uu}.

It must be emphasized that the $B_{\mu}=0$ No-Scale
boundary condition is the central agent affording this determination, as it is the extraction of the parameterized
parabolic curve of solutions in the two compensating variables which allows for a localized, bound nadir point to be
isolated by the Super No-Scale condition, dynamically determining {\it both} parameters.  The background surface of
$V_{\rm min}$ for the full parameter space outside the viable $B_{\mu}=0$ subset is, in contrast, a steadily inclined
and uninteresting function.  In our prior study, the local {\it minimum minimorum} of $V_{\rm min}$ for
selected inputs of $M_{V}$ and $m_{t}$ was taken to dynamically establish the values of  $M_{1/2}$ and $\tan \beta$.
Although $M_{1/2}$ and $\tan \beta$ have no {\it directly} established experimental values, they are severely indirectly constrained by
phenomenology in the context of this model~\cite{Li:2010ws,Li:2010mi}.  It is highly non-trivial that there should be
a strong accord between the top-down and bottom-up perspectives, but this is indeed precisely what has been observed~\cite{Li:2010uu}.

\section{\fsu5 SUSY Multijets at the $\sqrt{s} = 7$ TeV LHC}

The \fsu5 model space is bounded primarily by a set of ``bare-minimal'' experimental constraints distinguished 
by a great longevity of relevance, as defined in Ref.~\cite{Li:2011xu}. These include the top quark mass
$172.2~{\rm GeV} \leq m_{\rm t} \leq 174.4~{\rm GeV}$,
7-year WMAP cold dark matter relic density $0.1088 \leq \Omega_{\rm CDM}h^2 \leq
0.1158$~\cite{Komatsu:2010fb}, and precision LEP constraints on the SUSY mass content. We further
append to this classification an adherence to the defining high-scale boundary conditions of the model.
In light of recent developments, the favored parameter space may be further
circumscribed by the demands of a 124--126 GeV Higgs boson mass.  The surviving region is comprised
of a narrow strip of space confined to 400 $\le M_{1/2} \le$ 900 GeV, 19.4 $\le$ tan$\beta$ $\le$ 23, and 950
$\le M_V \le$ 6000 GeV. The border at the minimum $M_{1/2}$ =
400 GeV is required by the LEP constraints, while the maximum boundary at $M_{1/2}$ = 900 GeV prevents a
charged stau LSP at around tan$\beta \cong$ 23. In the bulk of the model space the lightness of the stau is leveraged to facilitate an appropriate dark matter relic density via stau-neutralino coannihilation.

The convergence of the predicted \fsu5 Higgs mass with the collider measured value
is achieved through contributions to the lightest CP-even Higgs boson mass from the flippons, calculated from the
RGE improved one-loop effective Higgs potential approach~\cite{Babu:2008ge,Martin:2009bg}. The
mechanism for the serendipitous mass shift is a pair of Yukawa interaction terms between the Higgs and
vector-like flippons in the superpotential, resulting in a 3--4 GeV upward shift in the Higgs mass to the
experimentally measured range~\cite{Li:2011ab}. Using the relevant shift in the Higgs mass-square
as approximated in Refs.~\cite{Huo:2011zt,Li:2011ab}, which implements a leading dependence of the
flippon mass $M_V$, larger shifts correspond to lighter vector-like flippons. This flippon induced
mechanism operates in synthesis with the top quark mass, whose elevation similarly raises the
non-flippon contributed Higgs mass. The cumulative result is a very narrow strip of model space, with the lower strip boundary truncated by the upper top quark mass extremity, and the upper strip boundary situated at the minimum Higgs mass of 124 GeV, conveniently
establishing a stable, thin band of experimentally viable points with which to explore new physics.

\begin{figure*}[htp]
        \centering
        \includegraphics[width=1.00\textwidth]{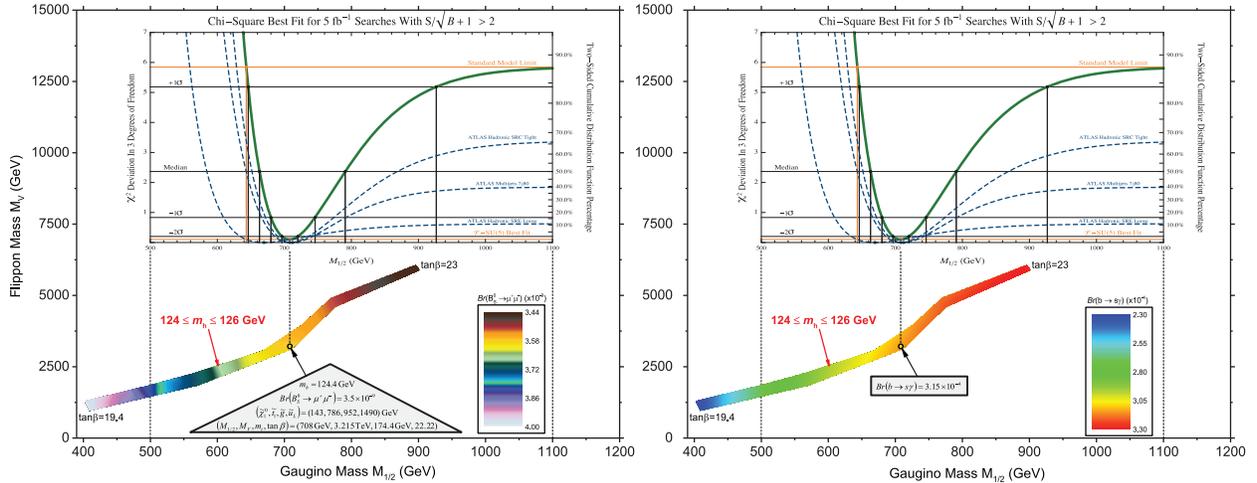}
        \caption{We depict the experimentally viable parameter space of No-Scale \fsu5 as a function of the gaugino mass $M_{1/2}$
and flippon mass $M_V$. The surviving model space after application of the bare-minimal constraints of
Ref.~\cite{Li:2011xu}
and Higgs boson mass calculations of Ref.~\cite{Li:2011ab} is illustrated by the narrow strip with the
smoothly contoured
color gradient. The gradient represents the total branching ratio (SM+SUSY) of the B-decay process \bs0
(left), and the total branching ratio (SM+SUSY) of $b \to s \gamma$ (right). The inset diagrams (with
linked horizontal scale) are the multi-axis cumulative \x2
fitting of Ref.~\cite{Li:2012tr}, depicting the best SUSY mass fit and Standard Model limit of only those
ATLAS and CMS
SUSY searches exhibiting a signal significance of \sb $>$ 2. The best fit benchmark of
Ref.~\cite{Li:2012tr} is highlighted at $M_{1/2}$ = 708 GeV, with $m_h$ = 124.4 GeV.}
        \label{fig:sliver}
\end{figure*}

The same flippon induced perturbation to the RGE unification structure of \fsu5 that was responsible
for facilitating a consistent application of the No-Scale boundary conditions near the Planck mass
also produces a key phenomenological signature.  The flat RGE evolution of the $SU(3)_C$ gaugino mass $M_3$, which 
mirrors the flatness of the $\beta$-coefficient $b_3 = 0$, suppresses the standard logarithmic mass enhancement at low-energy
and yields a SUSY spectrum $M(\widetilde{t}_1) < M(\widetilde{g}) < M(\widetilde{q})$
where the light stop $\widetilde{t}_1$ and gluino $\widetilde{g}$ are both less massive than all other squarks.
This highly unusual hierarchy produces a distinct event topology initiated by the pair-production
of heavy first or second generation squarks $\widetilde{q}$ and/or gluinos in the hard scattering
process, with the heavy squark likely to yield a quark-gluino pair $\widetilde{q} \rightarrow q \widetilde{g}$.
The gluino then has only two main channels available in the cascade decay,
$\widetilde{g} \rightarrow \widetilde{t}_1 \overline{t}$ or $\widetilde{g} \rightarrow q \overline{q} \widetilde{\chi}_1^0$,
with $\widetilde{t}_1 \rightarrow t \widetilde{\chi}_1^0$ or $\widetilde{t}_1 \rightarrow b \widetilde{\chi}_1^{\pm}$.
As $M_{1/2}$ increases, the stop-top channel becomes
dominant, ultimately reaching 100\% for $M_{1/2} \ge 729$ GeV. For $M_{1/2} < 729$ GeV, both avenues have
sufficient branching fractions to produce observable events at the LHC. Each gluino produces 2--6
hadronic jets, with the maximum of six jets realized in the gluino-mediated stop decay, so that a single gluino-gluino
pair-production event can net 4--12 jets.  After further fragmentation processes, the final event is
characterized by a definitive SUSY signal of high-multiplicity jets.

The most robust test of any supersymmetric model is the prediction of a unique signature plainly accounting
for observed anomalies in collider data.  The exceptional mass ordering in No-Scale \fsu5 provides a
distinctive marker at the LHC, since multijet events are expected to dominate a probed \fsu5 framework. We
first suggested in March 2011~\cite{Maxin:2011hy,Li:2011hr} that SUSY in an \fsu5 universe would become
manifest at the colliders in high-multiplicity jet events, extending this initial study in
Refs.~\cite{Li:2011rp,Li:2011fu,Li:2011xg,Li:2011av,Li:2011ab,Li:2012hm,Li:2012tr,Li:2012ix}.
The first ample accumulation of multijet data was released by the collaborations later in 2011 in
Refs.~\cite{PAS-SUS-11-003,Aad:2011ib,Aad:2011qa}, based upon 1 \fb of luminosity. Though the
number of events remaining after the collaboration data cuts was less than ten, there did
appear small but curious excesses beyond the SM estimates in these searches targeting multijet events.
The most prominent examples came from ATLAS, where the 7j80 ($\ge$ 7 jets and jet $p_T >$ 80 GeV) search of
Ref.~\cite{Aad:2011qa} and High Mass ($\ge$ 4 jets and jet $p_T >$ 80 GeV) search of Ref.~\cite{Aad:2011ib}
displayed interesting event production over the data-driven background estimates. Employing the signal significance
metric \sb, we computed a value of 1.1 for 7j80 and 1.3 for the High Mass search.  Despite the weak signal,
reasonably attributable to statistical fluctuations, No-Scale \fsu5 provided a neat and efficient explanation
for the minor over-productions in these two searches.  Despite the long odds at that time, those clean
fits prompted us to extrapolate from the ATLAS published statistics of Refs.~\cite{Aad:2011ib,Aad:2011qa}
to predict signal strengths of \sb = 1.9 for 7j80 and \sb = 3.0 for the High Mass~\cite{Li:2012hm} search
in the forthcoming 5 \fb data set at 7 TeV, assuming a legitimate physics origin for the intriguing over-production.

We provided a detailed analysis of the ATLAS and CMS 5 \fb observations at the 7 TeV LHC in
Ref.~\cite{Li:2012tr}, focused on those search strategies where the signal significance was strongest
and the largest number of events had accumulated, imposing \sb $>$ 2.0 as a minimal boundary. 
Strikingly, the 7j80~\cite{ATLAS-CONF-2012-037} search and the composite successors to the High Mass search were the only 5
\fb strategies to surmount this significance hurdle.
To elaborate, ATLAS essentially segregated the former High Mass $\ge$4 jet SUSY search of Ref.~\cite{Aad:2011ib} into three separate
searches of 4 jets, 5 jets, and 6 jets for the latter study, intended to isolate the $\widetilde{g}\widetilde{g}$,
$\widetilde{q}\widetilde{g}$, and $\widetilde{q}\widetilde{q}$ 0-lepton channels via $\widetilde{q}
\rightarrow q\widetilde{g}$ and $\widetilde{g} \rightarrow q
\overline{q} \widetilde{\chi}_1^0$~\cite{ATLAS-CONF-2012-033}. In addition to the ATLAS
7j80~\cite{ATLAS-CONF-2012-037}, these ATLAS 4-jet and 6-jet searches of
Ref.~\cite{ATLAS-CONF-2012-033}, referred to as SRC Tight and SRE Loose, respectively, were the only
other 5 \fb searches to achieve \sb $>$ 2.0 in all the ATLAS and CMS 5 \fb studies analyzed at that time.
Granting that the 1 \fb data sample is a subset of the 5 \fb data,
the signal strength nevertheless expanded in the precise proportionality expected.
The final 5 \fb 7 TeV ATLAS observations computed signal significances of \sb =
2.1 for 7j80~\cite{ATLAS-CONF-2012-037}, \sb = 3.2 for SRC Tight (4j)~\cite{ATLAS-CONF-2012-033},
and \sb = 2.6 for SRE Loose (6j)~\cite{ATLAS-CONF-2012-033}, in line with our predictions and very
consistent with the signal growth expected to be observed in an \fsu5 universe.

This enlarged signal strength simultaneously presented a golden opportunity to derive a best fit SUSY mass to the 5 \fb data
through a \x2 fitting procedure.  We demonstrated~\cite{Li:2012tr} clear internal consistency in the \fsu5 mass scale favored
by the various search windows, in addition to the described correlation across time in the signal growth.
This analysis favored sparticle masses of $m_{\widetilde{\chi}_1^0}$ = 143 GeV,
$m_{\widetilde{t}_1}$ = 786 GeV, $m_{\widetilde{g}}$ = 952 GeV, and $m_{\widetilde{u}_L}$ = 1490 GeV,
complementing a Higgs mass of $m_{h}$ = 124.4 GeV at the $M_{1/2}$ = 708 GeV well of the 5 \fb multi-axis
cumulative \x2 curve, combining the 7j80, SRC Tight, and SRE Loose search channels.
To exemplify this best fit at the \x2 minimum, we chose an $M_{1/2}$ = 708 GeV point as our standing favored
benchmark~\cite{Li:2012tr}.  The superimposed cumulative \x2
curve of Ref.~\cite{Li:2012tr} visibly showcases how the ATLAS 7j80, SRC Tight, and SRE Loose over-productive
search strategies illuminate the \fsu5 model space as naturally conforming to the collider observations.
By lowering the minimum threshold for signal significance to \sb $>$ 1.0, the CMS 5 \fb MT2 search strategy~\cite{:2012jx}
was included into our 5 \fb multi-axis \x2 fitting in Ref.~\cite{Li:2012ix} along with an additional ATLAS search, namely the 8j55
case from Ref.~\cite{ATLAS-CONF-2012-037}.  It was demonstrated in this manner that further non-trivial correlations exist between
the mass scale favored by independently productive ATLAS and CMS SUSY searches, bolstering the case against attribution 
of the excesses to random statistical fluctuations.

The data observations for the ATLAS multijet searches discussed here have shown a very
natural progression from 1 \fb to 5 ${\rm fb^{-1}}$. In fact, the \sb $\sim$ 3 signal significance of the
combined ATLAS 5 \fb multijet searches, which we can consider to be about 3$\sigma$, is near the same signal
level as the Higgs boson after 5 \fb at 7 TeV. With the Higgs boson now at the discovery threshold of 5$\sigma$
in the first 8 TeV data tranche, it would only be fitting if the ATLAS multijet SUSY searches
continued to track the Higgs signal strength. Looking forward and preparing for potentially more significant SUSY production as we shift to forthcoming larger LHC beam collision energies and hence greater numbers of statistics, we transition here to a more appropriate metric for measuring signal strength in the presence of larger excess event production beyond expectations, $2 \times (\sqrt{S+B} - \sqrt{B})$. We employ the background statistics derived by the ATLAS Collaboration for 5 \fb at 7 TeV from Ref.~\cite{ATLAS-CONF-2012-037}, though to determine an estimate of the SM background for 8 TeV, we scale up these ATLAS statistics using the same factor observed in our Monte Carlo for \fsu5 simulations. This estimator, while serving our limited scope here satisfactorily, can only be as reliable as the expectation of statistical, dynamic and procedural stability across the transition in energy, luminosity and model. We further assume here a static data cutting strategy between the ATLAS 7 and 8 TeV multijet searches. We indeed project that there should be a visible multijet signal strength sufficient for SUSY discovery in the isolated 15 \fb 8 TeV data, expected to be recorded in 2012 and processed in 2013, if the existing signal in the 5 \fb 7 TeV data is legitimately and wholly attributable to new physics. More precisely, assuming no important modifications to the background calibration procedures by ATLAS, we can project the 7j80 SUSY search tactics of Ref.~\cite{ATLAS-CONF-2012-037} to yield a signal significance of $2 \times (\sqrt{S+B} - \sqrt{B}) \sim 6$ for 15 \fb at 8 TeV, and $2 \times (\sqrt{S+B} - \sqrt{B}) \sim$ 7--8 for the SRC-Tight and SRE-Loose search strategies of Ref.~\cite{ATLAS-CONF-2012-033}. Although potentially quite susceptible to large statistical fluctuation, these rather strong signal projections nonetheless indicate that a probing of the \fsu5 framework at the LHC could indeed yield further tantalizing, and possibly convincing, evidence that nature herself is fundamentally supersymmetric. The summation of the 5 \fb of 7 TeV data to the 8 TeV data only improves the signal significance modestly. Moreover, the presence of excess events in the 15 \fb ATLAS multijet searches at 8 TeV will resoundingly indicate that random background anomalies are not the source of the 7 TeV multijet over-production. We find the predictable evolution of our SUSY exploration from the initial 1 \fb at 7 TeV to the 5 \fb at 7 TeV to warrant such positive speculation as we move forward to the 15 \fb at 8 TeV.

\section{Primordial Synthesis}

We now seek to synthesize the strip of model space supporting an $m_h \sim 125$ GeV Higgs boson~\cite{Li:2012yd,Li:2012qv,Li:2012jf} with the amalgamation of complementary supersymmetry experiments, including our 8 TeV conclusions of Ref.~\cite{Li:2012mr}. We begin with the original components
of our Golden Strip~\cite{Li:2010mi,Li:2011xu,Li:2011xg}, which are the key rare process limits on
\textit{Br}($b \to s \gamma$), \textit{Br}(\bs0), and $\Delta a_{\mu}$ on $(g-2)_{\mu}$ of the muon.
For $b \to s \gamma$, we use the latest world average of the Heavy Flavor Averaging Group (HFAG), BABAR,
Belle, and CLEO, which is $(3.55 \pm 0.24_{\rm exp} \pm 0.09_{\rm model}) \times
10^{-4}$~\cite{Barberio:2007cr}. An alternate approach to the average~\cite{Artuso:2009jw} yields
a slightly smaller central value, but also a lower error, suggesting
\textit{Br}$(b \to s\gamma) = (3.50 \pm 0.14_{\rm exp} \pm 0.10_{\rm model}) \times 10^{-4}$. See Ref.~\cite{Misiak:2010dz} for recent
discussion and analysis. The theoretical SM contribution at the next-to-next-to-leading order (NNLO)
is estimated at \textit{Br}$(b \to s \gamma) = (3.15 \pm 0.23) \times 10^{-4}$~\cite{Misiak:2006zs} and
\textit{Br}$(b \to s \gamma) = (2.98 \pm 0.26) \times 10^{-4}$~\cite{Becher:2006pu}. The addition of
these errors in quadrature provides the $2\sigma$ limits of
$2.86 \times 10^{-4} \le Br(b \to s \gamma) \le 4.24 \times 10^{-4}$. The recent precision improved LHCb constraints on the B-decay process \bs0 of
\textit{Br}(\bs0) $< 4.5(3.8) \times 10^{-9}$ at the 95\% (90\%) confidence level~\cite{Aaij:2012ac}
are employed here, though we find the entire viable \fsu5 parameter space lies comfortably below this
upper limit~\cite{Li:2012yd}. The new calculations of the tenth-order QED terms for the theoretical
prediction of $(g-2)_{\mu}$ engenders a favorable shift in $\Delta a_{\mu}$ in the context of \fsu5,
where we apply the $2\sigma$ uncertainty of $6.6 \times 10^{-10} \le \Delta a_{\mu} \le 41.4 \times
10^{-10}$. The $b \to s \gamma$ and $(g-2)_{\mu}$ effects reside at their lower boundaries in the 125 GeV
Higgs boson strip, as they exert pressure in opposing directions on $M_{1/2}$ since the leading gaugino and
squark contributions to \textit{Br}($b \to s \gamma$) admit an opposite sign to the Standard Model term
and Higgs contribution. On the contrary, the effect is additive for the non-Standard Model contribution
to $\Delta a_{\mu}$, establishing an upper limit on $M_{1/2}$. The SUSY contribution to
\textit{Br}($b \to s \gamma$) cannot be excessively large such that the Standard Model effect becomes minimized, thus
necessitating a sufficiently large, or lower bounded, $M_{1/2}$.

\begin{figure*}[htp]
        \centering
        \includegraphics[width=0.77\textwidth]{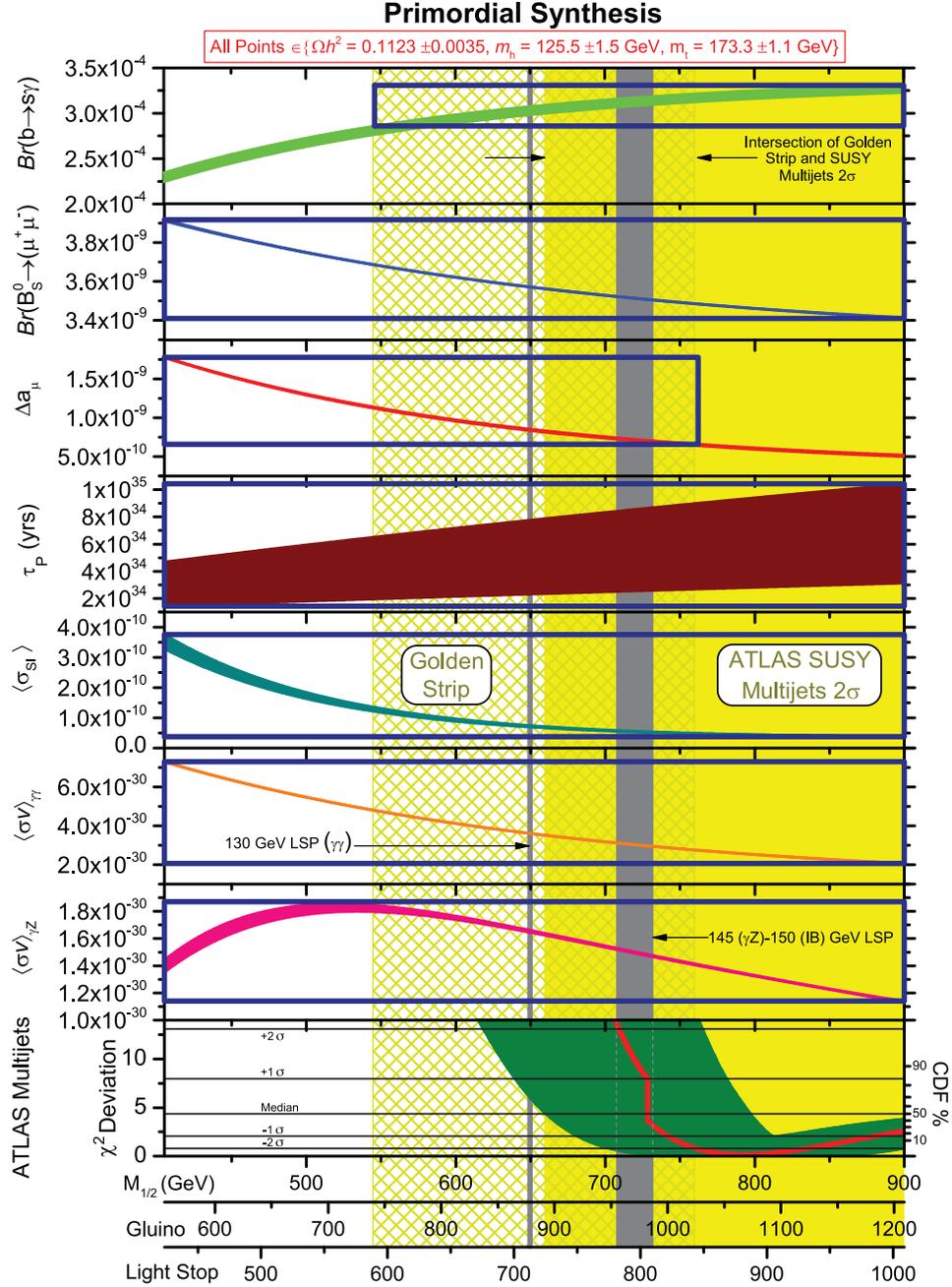}
        \caption{Primordial Synthesis of all currently progressing experiments searching for physics beyond the
Standard Model. All points depicted on each curve satisfy the conditions $0.1088 \le \Omega h^2 \le 0.1158$,
$124 \le m_h \le 127$ GeV, and $172.2 \le m_t \le 174.4$ GeV. Each curve thickness represents an uncertainty on
the strong coupling constant $0.1145 \le \alpha_s(M_Z) \le 0.1172$ (excluding the $\chi^2$ pane). The multi-axis
$\chi^2$ deviation in the bottom pane comprises an uncertainty derived from an increase and decrease by a factor
of 2 around the $\chi^2$ computed on the nominal number of \fsu5 events surviving all cuts (nominal value shown in
center of shaded curve).}
        \label{fig:primordial}
\end{figure*}

The computation of the rare-decay processes for all points in the 125 GeV Higgs boson strip are illustrated in
Figure (\ref{fig:primordial}). We implement a range on the strong coupling constant of $0.1145 \le \alpha_s(M_Z) \le 0.1172$
that tightly envelopes the central value of $\alpha_s(M_Z) = 0.1161$ that is
supported by recent direct observations~\cite{Bandurin:2011sh}, introducing a modest uncertainty
onto the calculation of each curve in Figure (\ref{fig:primordial}), represented by the contour
thickness in each pane. All SUSY particle masses, Higgs boson masses, relic densities, and constraints are computed with
{\tt MicrOMEGAs 2.4}~\cite{Belanger:2010gh}, applying the proprietary modification of the {\tt SuSpect 2.34}~\cite{Djouadi:2002ze}
codebase to run the flippon-enhanced RGEs. In Figure (\ref{fig:primordial}), the boxed curve segments depict the experimentally observed
$2\sigma$ values.

We now expand our original Golden Strip to encompass proton decay and dark matter detection experiments.
The $p\rightarrow e^+ \pi^0$ mode in \fsu5 is depicted in Figure (\ref{fig:primordial}), indicative of
the large pervasive uncertainty propagated into the proton lifetime from the large QCD uncertainties in
$\alpha_s(M_Z)$. We apply the Super-Kamiokande established lower bound of $1.4 \times 10^{34}$ years at
the 90\% confidence level for the partial lifetime in the $p\rightarrow e^+ \pi^0$
mode~\cite{Hewett:2012ns}. For the spin-independent dark matter-nucleon cross section, the XENON100
experiment has probed down to $2\times 10^{-9}~{\rm pb}$ ($2\times 10^{-45}~{\rm cm^2}$) for a WIMP mass
of 55 GeV~\cite{Aprile:2012nq}, also at the 90\% confidence level. The No-Scale \fsu5 viable model space
shown in Figure (\ref{fig:primordial}) lies entirely below this upper bound~\cite{Li:2011in}.

The observation of a 130 GeV gamma-ray line with annihilation cross-section
$\left\langle \sigma v \right\rangle \sim 10^{-27}~{\rm cm^3/s}$~\cite{Weniger:2012tx} emanating from our galactic center by
the FERMI-LAT Space Telescope has initiated investigations into whether such a monochromatic line could
be attributed to dark matter annihilations, an argument amplified by the lack of any known astrophysical
source capable of producing a tantamount signature. The lightest neutralino mass at the minimum of the \x2
fit to the ATLAS multijet and light stop excesses is $m_{\chi} = 145$ GeV, clearly highlighted as the
benchmark model in Figure (\ref{fig:sliver}). Conjecturing the observed photon line originates from
neutralino annihilations into a $Z$-boson and gamma-ray via $\widetilde{\chi} \widetilde{\chi} \to Z
\gamma$, we can compute the kinematics for a non-relativistic lightest neutralino
$\widetilde{\chi}_1^0$ as
\begin{eqnarray}
E_{\gamma} = M_{\chi} - \frac{M_Z^2}{4 M_{\chi}} \, ,
\label{eq:E}
\end{eqnarray}
which gives
\begin{eqnarray}
M_{\chi} = \frac{E_{\gamma}}{2}\left[1 + \sqrt{1 + \left(\frac{M_Z}{E_{\gamma}}\right)^2}\right]
\label{eq:M}
\end{eqnarray}
Using $E_{\gamma} \simeq 130$ GeV and $M_Z = 91.187$ GeV, we arrive at
\begin{eqnarray}
M_{\chi} = 144.4~ {\rm GeV} \, ,
\nonumber
\label{eq:W}
\end{eqnarray}
which is consistent with the No-Scale \fsu5 lightest neutralino mass of $M_{\chi} = 145$ GeV
at the best fit to the multijet and light stop excesses at the LHC $and$ generates an $m_h \simeq 125.5$ GeV Higgs boson mass. The fit is near $m_{\chi} \sim 150$ GeV for internal bremsstrahlung~\cite{Bringmann:2012vr}. We allow for the potential combination of all of the
above that could land a WIMP mass somewhere in the range $130 \lesssim m_{\chi} \lesssim 150$ GeV, and as a
consequence, we annotate the 130-150 GeV LSP mass region in Figure (\ref{fig:primordial}). The most
recent FERMI-LAT Collaboration upper bound on the gamma-ray annihilation cross section is
$\left\langle \sigma v \right\rangle \sim 10^{-26}~{\rm cm^3/s}$~\cite{Ackermann:2012qk}, which we
use in Figure (\ref{fig:primordial}), allowing for a possible boost factor, as elaborated subsequently.

The \fsu5 lightest neutralino composition is greater than 99\% bino, therefore, we cannot neglect the
quite small $\widetilde{\chi} \widetilde{\chi} \to Z \gamma$ annihilation cross-section
$\left\langle \sigma v \right\rangle \sim 10^{-30}~{\rm cm^3/s}$, about three orders of magnitude less
than the FERMI-LAT telescope observations. On the other hand, a recent analysis~\cite{Hektor:2012kc}
of extra-galactic clusters uncovering synonymous 130 GeV gamma-ray lines has determined an appropriate
subhalo boost factor in this context of $\sim 1000$ relative to the galactic center.  We do not consider it
implausible that an overall unaccounted boost factor of similar magnitude might reconcile this apparent discrepancy
of scale.  For now, we are content to simply make note of the interesting correlation that exists between 
145 GeV \fsu5 neutralino annihilations and the unexplained 130 GeV gamma-ray line observed by
the FERMI-LAT space telescope, irrespective of the absolute $\left\langle \sigma v \right\rangle$ 
cross-section magnitude. 

We include in Figure (\ref{fig:primordial}) the multi-axis $\chi^2$ of Ref.~\cite{Li:2012mr},
computed from those 8 TeV ATLAS multijet searches that display evidence of over-production above
background expectations. The vertical yellow band in Figure (\ref{fig:primordial}) depicts the
$2\sigma$ range around the $\chi^2$ minimum computed from the nominal number of \fsu5 simulated events
times 0.50, bordered by the lower $2\sigma$ boundary at about $M_{1/2} \sim 660$ GeV. The Golden Strip is
represented by the cross-hatched region, confined by the lower $2\sigma$ boundary on
\textit{Br}$(b \to s \gamma)$ at its lower $M_{1/2} \sim 545$ GeV limit, and by the lower $2\sigma$ boundary on
$\Delta a_{\mu}$ at the Golden Strip's upper $M_{1/2} \sim 760$ GeV limit. Demonstrated in Figure
(\ref{fig:primordial}) is the intersection of these two bands of model space defined by the
$2\sigma$ observable regions of completely uncorrelated experiments, though apparently exhibiting interesting
evidence of correlated behavior in a No-Scale \fsu5 framework. To further heighten the intrigue, the
130-150 GeV LSP model space corresponding to the FERMI-LAT Space Telescope observations of a 130 GeV
monochromatic gamma-ray line from the galactic center also very curiously lies snugly within the
intersection of all experiments. Notice that the gluino and light stop mass scales are inserted at the
bottom of Figure (\ref{fig:primordial}). Due to the characteristic rescaling property of No-Scale
\fsu5, a direct proportional relationship exists between the SUSY spectrum and gaugino $M_{1/2}$,
permitting a simple visual inspection of the associated gluino and light stop masses for any specified
$M_{1/2}$.

It is worth emphasizing again that all points delineated by the curves in each pane in Figure
(\ref{fig:primordial}) are themselves the intersection of three critical parameters measured to high
precision in current experiments, namely the 7-year WMAP relic density $0.1088 \le \Omega h^2 \le 0.1158$,
a 124-127 GeV light Higgs boson mass, and a $172.2 \le m_t \le 174.4$ GeV top quark mass. Therefore, at the
present time, we can find no experiment pertinent to the supersymmetric parameter space that is not in
conformance with the narrow band of No-Scale \fsu5 model space from $660 \lesssim M_{1/2} \lesssim 760$
GeV, which corresponds to sparticle masses of $133 \lesssim M({\rm \widetilde{\chi}_1^0}) \lesssim
160$ GeV, $725 \lesssim M({\rm \widetilde{t}_1}) \lesssim 845$ GeV, and
$890 \lesssim M({\rm \widetilde{g}}) \lesssim 1025$ GeV. Such a mutual interrelation between
all relevant experiments seems to strongly belie attribution to random stochastics.

The proximity of the 145-150 GeV LSP strip that resides within the theoretically and phenomenologically
favored \fsu5 parameter space defined by all model constraints, in relation to the minimum of our
multi-axis $\chi^2$ curve, recalls to mind a very similar level of statistical adjacency shared by the 
updated $\chi^2$ curves for the experimental Higgs boson mass measurements ($m_h \sim 125$ GeV) with 
the mass region theoretically and phenomenologically favored by electroweak precision measurements
at $m_h = 94^{+29}_{-24}$ GeV~\cite{LEPEWWG}.  The difference of about one standard deviation between the
empirically measured Higgs boson mass and the electroweak precision favored region is roughly akin to the
statistical margin separating the LHC SUSY multijet measurements and the optimum phenomenological \fsu5 region, where
we would assign a standard fluctuation width of about 60~GeV to
deviations in the downward mass direction, and 200 GeV to the upper $\chi^2$ median intersection. Thus, we may take great satisfaction that such a level
of consistency is displayed between experiment and theory in \fsu5, supported by relevant historical precedent.

As two points of potentially relevant interest, we must also remark in passing on recent
developments regarding the measurement of the top quark mass and the strong coupling constant.
An external study based on ATLAS inclusive jet cross section data~\cite{Malaescu:2012iq} has suggested
the value $\alpha_s(M_Z) = 0.1151$, which is slightly lower than the world central value of 0.1161
on which we above remarked. Also, the CMS Collaboration has recently announced~\cite{:2012cz}
the world's single most precise top quark mass measurement at $m_h = 173.49 \pm 1.07$~GeV, with
a central value slightly above the existing world average. Moreover, the latest measurements by ATLAS show central values of $m_t = 174.5$
GeV~\cite{ATLAS:2012aj}, $m_t = 174.9$ GeV~\cite{ATLAS-CONF-2012-030}, and $m_t = 175.2$
GeV~\cite{ATLAS-CONF-2012-082}, all modestly elevated above the world average central value. In Ref.~\cite{Li:2012jf}, we investigated on the roles that a slightly elevated top quark mass, and a slightly reduced strong coupling could play in facilitating
satisfaction of the central Higgs mass measurements in the range of $125$--$126$~GeV, without
resorting to an overly heavy squark spectrum or extremities in the error margins for the Higgs mass itself.
The lowering of $\alpha_s$ while maintaining consistency with precision electroweak scale data
is an accommodation to which the flipped $SU(5)$ GUT is particularly well historically adapted~\cite{Ellis:1995at}.
An interesting side effect of this modification is an escalation in the proton decay rate linked 
to a parallel reduction in the GUT scale $M_{32}$.

We close our discussion of Figure (\ref{fig:primordial}) by remarking on the striking familiarity of
this figure to the correlation of predicted and observed light elemental abundances with the value of the
baryon-to-photon ratio given by the observations of the Cosmic Microwave Background (CMB) by WMAP. The
amazing consistency with which predictions of light element abundances by Primordial Nucleosynthesis
demonstrates with astronomical observations, while also compatible with the independently measured
CMB, provide powerful corroboration of the Big Bang Theory. We envision a compelling parallel here
amongst the synthesis of light elements predicted by Primordial Nucleosynthesis and observed by
experiments, with the synthesis in an ubiquitous \fsu5 structure in nature of all currently progressing
experiments searching for physics beyond the Standard Model, to which we aptly offer the description
\textit{Primordial Synthesis}. Analogous to the consistency encountered between theory and
experiment of light elemental abundances in Primordial Nucleosynthesis that provides a convincing
connection to the Big Bang Theory, we suggest that the consistency revealed in Figure
(\ref{fig:primordial}) between all the BSM experiments in No-Scale \fsu5 Primordial Synthesis
presents persuasive indications of BSM physics currently being probed at the LHC and indeed possibly all
the experiments involved in searching for the parameters in Figure (\ref{fig:primordial}).


\begin{acknowledgments}
This research was supported in part
by the DOE grant DE-FG03-95-Er-40917 (TL and DVN),
by the Natural Science Foundation of China
under grant numbers 10821504, 11075194, 11135003, and 11275246 (TL),
and by the Mitchell-Heep Chair in High Energy Physics (JAM).
We also thank Sam Houston State University
for providing high performance computing resources.
\end{acknowledgments}


\bibliography{bibliography}

\begin{thebibliography}{83}
\expandafter\ifx\csname natexlab\endcsname\relax\def\natexlab#1{#1}\fi
\expandafter\ifx\csname bibnamefont\endcsname\relax
  \def\bibnamefont#1{#1}\fi
\expandafter\ifx\csname bibfnamefont\endcsname\relax
  \def\bibfnamefont#1{#1}\fi
\expandafter\ifx\csname citenamefont\endcsname\relax
  \def\citenamefont#1{#1}\fi
\expandafter\ifx\csname url\endcsname\relax
  \def\url#1{\texttt{#1}}\fi
\expandafter\ifx\csname urlprefix\endcsname\relax\def\urlprefix{URL }\fi
\providecommand{\bibinfo}[2]{#2}
\providecommand{\eprint}[2][]{\url{#2}}

\bibitem[{\citenamefont{Li et~al.}(2011{\natexlab{a}})\citenamefont{Li, Maxin,
  Nanopoulos, and Walker}}]{Li:2010ws}
\bibinfo{author}{\bibfnamefont{T.}~\bibnamefont{Li}},
  \bibinfo{author}{\bibfnamefont{J.~A.} \bibnamefont{Maxin}},
  \bibinfo{author}{\bibfnamefont{D.~V.} \bibnamefont{Nanopoulos}},
  \bibnamefont{and} \bibinfo{author}{\bibfnamefont{J.~W.}
  \bibnamefont{Walker}}, {``}\bibinfo{title}{{The Golden Point of No-Scale and
  No-Parameter ${\cal F}$-$SU(5)$}},{''} \bibinfo{journal}{Phys. Rev.}
  \textbf{\bibinfo{volume}{D83}}, \bibinfo{pages}{056015}
  (\bibinfo{year}{2011}{\natexlab{a}}), \eprint{1007.5100}.

\bibitem[{\citenamefont{Li et~al.}(2011{\natexlab{b}})\citenamefont{Li, Maxin,
  Nanopoulos, and Walker}}]{Li:2010mi}
\bibinfo{author}{\bibfnamefont{T.}~\bibnamefont{Li}},
  \bibinfo{author}{\bibfnamefont{J.~A.} \bibnamefont{Maxin}},
  \bibinfo{author}{\bibfnamefont{D.~V.} \bibnamefont{Nanopoulos}},
  \bibnamefont{and} \bibinfo{author}{\bibfnamefont{J.~W.}
  \bibnamefont{Walker}}, {``}\bibinfo{title}{{The Golden Strip of Correlated
  Top Quark, Gaugino, and Vectorlike Mass In No-Scale, No-Parameter
  F-SU(5)}},{''} \bibinfo{journal}{Phys. Lett.}
  \textbf{\bibinfo{volume}{B699}}, \bibinfo{pages}{164}
  (\bibinfo{year}{2011}{\natexlab{b}}), \eprint{1009.2981}.

\bibitem[{\citenamefont{Barr}(1982)}]{Barr:1981qv}
\bibinfo{author}{\bibfnamefont{S.~M.} \bibnamefont{Barr}},
  {``}\bibinfo{title}{{A New Symmetry Breaking Pattern for $SO(10)$ and Proton
  Decay}},{''} \bibinfo{journal}{Phys. Lett.} \textbf{\bibinfo{volume}{B112}},
  \bibinfo{pages}{219} (\bibinfo{year}{1982}).

\bibitem[{\citenamefont{Derendinger et~al.}(1984)\citenamefont{Derendinger,
  Kim, and Nanopoulos}}]{Derendinger:1983aj}
\bibinfo{author}{\bibfnamefont{J.~P.} \bibnamefont{Derendinger}},
  \bibinfo{author}{\bibfnamefont{J.~E.} \bibnamefont{Kim}}, \bibnamefont{and}
  \bibinfo{author}{\bibfnamefont{D.~V.} \bibnamefont{Nanopoulos}},
  {``}\bibinfo{title}{{Anti-$SU(5)$}},{''} \bibinfo{journal}{Phys. Lett.}
  \textbf{\bibinfo{volume}{B139}}, \bibinfo{pages}{170} (\bibinfo{year}{1984}).

\bibitem[{\citenamefont{Antoniadis et~al.}(1987)\citenamefont{Antoniadis,
  Ellis, Hagelin, and Nanopoulos}}]{Antoniadis:1987dx}
\bibinfo{author}{\bibfnamefont{I.}~\bibnamefont{Antoniadis}},
  \bibinfo{author}{\bibfnamefont{J.~R.} \bibnamefont{Ellis}},
  \bibinfo{author}{\bibfnamefont{J.~S.} \bibnamefont{Hagelin}},
  \bibnamefont{and} \bibinfo{author}{\bibfnamefont{D.~V.}
  \bibnamefont{Nanopoulos}}, {``}\bibinfo{title}{{Supersymmetric Flipped
  $SU(5)$ Revitalized}},{''} \bibinfo{journal}{Phys. Lett.}
  \textbf{\bibinfo{volume}{B194}}, \bibinfo{pages}{231} (\bibinfo{year}{1987}).

\bibitem[{\citenamefont{Jiang et~al.}(2007)\citenamefont{Jiang, Li, and
  Nanopoulos}}]{Jiang:2006hf}
\bibinfo{author}{\bibfnamefont{J.}~\bibnamefont{Jiang}},
  \bibinfo{author}{\bibfnamefont{T.}~\bibnamefont{Li}}, \bibnamefont{and}
  \bibinfo{author}{\bibfnamefont{D.~V.} \bibnamefont{Nanopoulos}},
  {``}\bibinfo{title}{{Testable Flipped $SU(5) \times U(1)_X$ Models}},{''}
  \bibinfo{journal}{Nucl. Phys.} \textbf{\bibinfo{volume}{B772}},
  \bibinfo{pages}{49} (\bibinfo{year}{2007}), \eprint{hep-ph/0610054}.

\bibitem[{\citenamefont{Jiang et~al.}(2009)\citenamefont{Jiang, Li, Nanopoulos,
  and Xie}}]{Jiang:2009zza}
\bibinfo{author}{\bibfnamefont{J.}~\bibnamefont{Jiang}},
  \bibinfo{author}{\bibfnamefont{T.}~\bibnamefont{Li}},
  \bibinfo{author}{\bibfnamefont{D.~V.} \bibnamefont{Nanopoulos}},
  \bibnamefont{and} \bibinfo{author}{\bibfnamefont{D.}~\bibnamefont{Xie}},
  {``}\bibinfo{title}{{F-$SU(5)$}},{''} \bibinfo{journal}{Phys. Lett.}
  \textbf{\bibinfo{volume}{B677}}, \bibinfo{pages}{322} (\bibinfo{year}{2009}).

\bibitem[{\citenamefont{Jiang et~al.}(2010)\citenamefont{Jiang, Li, Nanopoulos,
  and Xie}}]{Jiang:2009za}
\bibinfo{author}{\bibfnamefont{J.}~\bibnamefont{Jiang}},
  \bibinfo{author}{\bibfnamefont{T.}~\bibnamefont{Li}},
  \bibinfo{author}{\bibfnamefont{D.~V.} \bibnamefont{Nanopoulos}},
  \bibnamefont{and} \bibinfo{author}{\bibfnamefont{D.}~\bibnamefont{Xie}},
  {``}\bibinfo{title}{{Flipped $SU(5) \times U(1)_X$ Models from
  F-Theory}},{''} \bibinfo{journal}{Nucl. Phys.}
  \textbf{\bibinfo{volume}{B830}}, \bibinfo{pages}{195} (\bibinfo{year}{2010}),
  \eprint{0905.3394}.

\bibitem[{\citenamefont{Li et~al.}(2011{\natexlab{c}})\citenamefont{Li,
  Nanopoulos, and Walker}}]{Li:2010dp}
\bibinfo{author}{\bibfnamefont{T.}~\bibnamefont{Li}},
  \bibinfo{author}{\bibfnamefont{D.~V.} \bibnamefont{Nanopoulos}},
  \bibnamefont{and} \bibinfo{author}{\bibfnamefont{J.~W.}
  \bibnamefont{Walker}}, {``}\bibinfo{title}{{Elements of F-ast Proton
  Decay}},{''} \bibinfo{journal}{Nucl. Phys.} \textbf{\bibinfo{volume}{B846}},
  \bibinfo{pages}{43} (\bibinfo{year}{2011}{\natexlab{c}}), \eprint{1003.2570}.

\bibitem[{\citenamefont{Li et~al.}(2011{\natexlab{d}})\citenamefont{Li, Maxin,
  Nanopoulos, and Walker}}]{Li:2010rz}
\bibinfo{author}{\bibfnamefont{T.}~\bibnamefont{Li}},
  \bibinfo{author}{\bibfnamefont{J.~A.} \bibnamefont{Maxin}},
  \bibinfo{author}{\bibfnamefont{D.~V.} \bibnamefont{Nanopoulos}},
  \bibnamefont{and} \bibinfo{author}{\bibfnamefont{J.~W.}
  \bibnamefont{Walker}}, {``}\bibinfo{title}{{Dark Matter, Proton Decay and
  Other Phenomenological Constraints in ${\cal F}$-SU(5)}},{''}
  \bibinfo{journal}{Nucl.Phys.} \textbf{\bibinfo{volume}{B848}},
  \bibinfo{pages}{314} (\bibinfo{year}{2011}{\natexlab{d}}),
  \eprint{1003.4186}.

\bibitem[{\citenamefont{Cremmer et~al.}(1983)\citenamefont{Cremmer, Ferrara,
  Kounnas, and Nanopoulos}}]{Cremmer:1983bf}
\bibinfo{author}{\bibfnamefont{E.}~\bibnamefont{Cremmer}},
  \bibinfo{author}{\bibfnamefont{S.}~\bibnamefont{Ferrara}},
  \bibinfo{author}{\bibfnamefont{C.}~\bibnamefont{Kounnas}}, \bibnamefont{and}
  \bibinfo{author}{\bibfnamefont{D.~V.} \bibnamefont{Nanopoulos}},
  {``}\bibinfo{title}{{Naturally Vanishing Cosmological Constant in $N=1$
  Supergravity}},{''} \bibinfo{journal}{Phys. Lett.}
  \textbf{\bibinfo{volume}{B133}}, \bibinfo{pages}{61} (\bibinfo{year}{1983}).

\bibitem[{\citenamefont{Ellis et~al.}(1984{\natexlab{a}})\citenamefont{Ellis,
  Lahanas, Nanopoulos, and Tamvakis}}]{Ellis:1983sf}
\bibinfo{author}{\bibfnamefont{J.~R.} \bibnamefont{Ellis}},
  \bibinfo{author}{\bibfnamefont{A.~B.} \bibnamefont{Lahanas}},
  \bibinfo{author}{\bibfnamefont{D.~V.} \bibnamefont{Nanopoulos}},
  \bibnamefont{and} \bibinfo{author}{\bibfnamefont{K.}~\bibnamefont{Tamvakis}},
  {``}\bibinfo{title}{{No-Scale Supersymmetric Standard Model}},{''}
  \bibinfo{journal}{Phys. Lett.} \textbf{\bibinfo{volume}{B134}},
  \bibinfo{pages}{429} (\bibinfo{year}{1984}{\natexlab{a}}).

\bibitem[{\citenamefont{Ellis et~al.}(1984{\natexlab{b}})\citenamefont{Ellis,
  Kounnas, and Nanopoulos}}]{Ellis:1983ei}
\bibinfo{author}{\bibfnamefont{J.~R.} \bibnamefont{Ellis}},
  \bibinfo{author}{\bibfnamefont{C.}~\bibnamefont{Kounnas}}, \bibnamefont{and}
  \bibinfo{author}{\bibfnamefont{D.~V.} \bibnamefont{Nanopoulos}},
  {``}\bibinfo{title}{{Phenomenological $SU(1,1)$ Supergravity}},{''}
  \bibinfo{journal}{Nucl. Phys.} \textbf{\bibinfo{volume}{B241}},
  \bibinfo{pages}{406} (\bibinfo{year}{1984}{\natexlab{b}}).

\bibitem[{\citenamefont{Ellis et~al.}(1984{\natexlab{c}})\citenamefont{Ellis,
  Kounnas, and Nanopoulos}}]{Ellis:1984bm}
\bibinfo{author}{\bibfnamefont{J.~R.} \bibnamefont{Ellis}},
  \bibinfo{author}{\bibfnamefont{C.}~\bibnamefont{Kounnas}}, \bibnamefont{and}
  \bibinfo{author}{\bibfnamefont{D.~V.} \bibnamefont{Nanopoulos}},
  {``}\bibinfo{title}{{No Scale Supersymmetric Guts}},{''}
  \bibinfo{journal}{Nucl. Phys.} \textbf{\bibinfo{volume}{B247}},
  \bibinfo{pages}{373} (\bibinfo{year}{1984}{\natexlab{c}}).

\bibitem[{\citenamefont{Lahanas and Nanopoulos}(1987)}]{Lahanas:1986uc}
\bibinfo{author}{\bibfnamefont{A.~B.} \bibnamefont{Lahanas}} \bibnamefont{and}
  \bibinfo{author}{\bibfnamefont{D.~V.} \bibnamefont{Nanopoulos}},
  {``}\bibinfo{title}{{The Road to No Scale Supergravity}},{''}
  \bibinfo{journal}{Phys. Rept.} \textbf{\bibinfo{volume}{145}},
  \bibinfo{pages}{1} (\bibinfo{year}{1987}).

\bibitem[{\citenamefont{Bennett et~al.}(2004)}]{Bennett:2004pv}
\bibinfo{author}{\bibfnamefont{G.~W.} \bibnamefont{Bennett}}
  \bibnamefont{et~al.} (\bibinfo{collaboration}{Muon g-2}),
  {``}\bibinfo{title}{{Measurement of the negative muon anomalous magnetic
  moment to 0.7-ppm}},{''} \bibinfo{journal}{Phys. Rev. Lett.}
  \textbf{\bibinfo{volume}{92}}, \bibinfo{pages}{161802}
  (\bibinfo{year}{2004}), \eprint{hep-ex/0401008}.

\bibitem[{\citenamefont{Barberio et~al.}(2007)}]{Barberio:2007cr}
\bibinfo{author}{\bibfnamefont{E.}~\bibnamefont{Barberio}} \bibnamefont{et~al.}
  (\bibinfo{collaboration}{Heavy Flavor Averaging Group (HFAG)}),
  {``}\bibinfo{title}{{Averages of $b-$hadron properties at the end of
  2006}},{''} (\bibinfo{year}{2007}), \eprint{0704.3575}.

\bibitem[{\citenamefont{Misiak et~al.}(2007)}]{Misiak:2006zs}
\bibinfo{author}{\bibfnamefont{M.}~\bibnamefont{Misiak}} \bibnamefont{et~al.},
  {``}\bibinfo{title}{{The first estimate of Br$(\overline{B} \rightarrow X_{s}
  \gamma)$ at ${\cal O}({\alpha}^{2}_{s})$}},{''} \bibinfo{journal}{Phys. Rev.
  Lett.} \textbf{\bibinfo{volume}{98}}, \bibinfo{pages}{022002}
  (\bibinfo{year}{2007}), \eprint{hep-ph/0609232}.

\bibitem[{\citenamefont{Komatsu et~al.}(2010)}]{Komatsu:2010fb}
\bibinfo{author}{\bibfnamefont{E.}~\bibnamefont{Komatsu}} \bibnamefont{et~al.}
  (\bibinfo{collaboration}{WMAP}), {``}\bibinfo{title}{{Seven-Year Wilkinson
  Microwave Anisotropy Probe (WMAP) Observations: Cosmological
  Interpretation}},{''} \bibinfo{journal}{Astrophys.J.Suppl.}
  \textbf{\bibinfo{volume}{192}}, \bibinfo{pages}{18} (\bibinfo{year}{2010}),
  \eprint{1001.4538}.

\bibitem[{:19(2010)}]{:1900yx}
{``}\bibinfo{title}{{Combination of CDF and D0 Results on the Mass of the Top
  Quark using up to 5.6 $fb^{-1}$ of data (The CDF and D0 Collaboration)}},{''}
  (\bibinfo{year}{2010}), \eprint{1007.3178}.

\bibitem[{\citenamefont{Li et~al.}(2011{\natexlab{e}})\citenamefont{Li, Maxin,
  Nanopoulos, and Walker}}]{Li:2010uu}
\bibinfo{author}{\bibfnamefont{T.}~\bibnamefont{Li}},
  \bibinfo{author}{\bibfnamefont{J.~A.} \bibnamefont{Maxin}},
  \bibinfo{author}{\bibfnamefont{D.~V.} \bibnamefont{Nanopoulos}},
  \bibnamefont{and} \bibinfo{author}{\bibfnamefont{J.~W.}
  \bibnamefont{Walker}}, {``}\bibinfo{title}{{Super No-Scale ${\cal
  F}$-$SU(5)$: Resolving the Gauge Hierarchy Problem by Dynamic Determination
  of $M_{1/2}$ and $\tan\beta$}},{''} \bibinfo{journal}{Phys. Lett. B}
  \textbf{\bibinfo{volume}{703}}, \bibinfo{pages}{469}
  (\bibinfo{year}{2011}{\natexlab{e}}), \eprint{1010.4550}.

\bibitem[{\citenamefont{Nakamura}(2003)}]{Nakamura:2003hk}
\bibinfo{author}{\bibfnamefont{K.}~\bibnamefont{Nakamura}},
  {``}\bibinfo{title}{{Hyper-Kamiokande: A next generation water Cherenkov
  detector}},{''} \bibinfo{journal}{Int. J. Mod. Phys.}
  \textbf{\bibinfo{volume}{A18}}, \bibinfo{pages}{4053} (\bibinfo{year}{2003}).

\bibitem[{\citenamefont{Raby et~al.}(2008)}]{Raby:2008pd}
\bibinfo{author}{\bibfnamefont{S.}~\bibnamefont{Raby}} \bibnamefont{et~al.},
  {``}\bibinfo{title}{{DUSEL Theory White Paper}},{''} (\bibinfo{year}{2008}),
  \eprint{0810.4551}.

\bibitem[{\citenamefont{Li et~al.}(2010)\citenamefont{Li, Nanopoulos, and
  Walker}}]{Li:2009fq}
\bibinfo{author}{\bibfnamefont{T.}~\bibnamefont{Li}},
  \bibinfo{author}{\bibfnamefont{D.~V.} \bibnamefont{Nanopoulos}},
  \bibnamefont{and} \bibinfo{author}{\bibfnamefont{J.~W.}
  \bibnamefont{Walker}}, {``}\bibinfo{title}{{Fast Proton Decay}},{''}
  \bibinfo{journal}{Phys. Lett.} \textbf{\bibinfo{volume}{B693}},
  \bibinfo{pages}{580} (\bibinfo{year}{2010}), \eprint{0910.0860}.

\bibitem[{\citenamefont{Huo et~al.}(2011)\citenamefont{Huo, Li, Nanopoulos, and
  Tong}}]{Huo:2011zt}
\bibinfo{author}{\bibfnamefont{Y.}~\bibnamefont{Huo}},
  \bibinfo{author}{\bibfnamefont{T.}~\bibnamefont{Li}},
  \bibinfo{author}{\bibfnamefont{D.~V.} \bibnamefont{Nanopoulos}},
  \bibnamefont{and} \bibinfo{author}{\bibfnamefont{C.}~\bibnamefont{Tong}},
  {``}\bibinfo{title}{{The Lightest CP-Even Higgs Boson Mass in the Testable
  Flipped $SU(5) \times U(1)_X$ Models from F-Theory}},{''}
  (\bibinfo{year}{2011}), \eprint{1109.2329}.

\bibitem[{\citenamefont{Kyae and Shafi}(2006)}]{Kyae:2005nv}
\bibinfo{author}{\bibfnamefont{B.}~\bibnamefont{Kyae}} \bibnamefont{and}
  \bibinfo{author}{\bibfnamefont{Q.}~\bibnamefont{Shafi}},
  {``}\bibinfo{title}{{Flipped SU(5) predicts delta(T)/T}},{''}
  \bibinfo{journal}{Phys. Lett.} \textbf{\bibinfo{volume}{B635}},
  \bibinfo{pages}{247} (\bibinfo{year}{2006}), \eprint{hep-ph/0510105}.

\bibitem[{\citenamefont{Ellis et~al.}(2002)\citenamefont{Ellis, Nanopoulos, and
  Walker}}]{Ellis:2002vk}
\bibinfo{author}{\bibfnamefont{J.~R.} \bibnamefont{Ellis}},
  \bibinfo{author}{\bibfnamefont{D.~V.} \bibnamefont{Nanopoulos}},
  \bibnamefont{and} \bibinfo{author}{\bibfnamefont{J.}~\bibnamefont{Walker}},
  {``}\bibinfo{title}{{Flipping SU(5) out of trouble}},{''}
  \bibinfo{journal}{Phys. Lett.} \textbf{\bibinfo{volume}{B550}},
  \bibinfo{pages}{99} (\bibinfo{year}{2002}), \eprint{hep-ph/0205336}.

\bibitem[{\citenamefont{Lopez et~al.}(1993{\natexlab{a}})\citenamefont{Lopez,
  Nanopoulos, and Yuan}}]{Lopez:1992kg}
\bibinfo{author}{\bibfnamefont{J.~L.} \bibnamefont{Lopez}},
  \bibinfo{author}{\bibfnamefont{D.~V.} \bibnamefont{Nanopoulos}},
  \bibnamefont{and} \bibinfo{author}{\bibfnamefont{K.-j.} \bibnamefont{Yuan}},
  {``}\bibinfo{title}{{The Search for a realistic flipped $SU(5)$ string
  model}},{''} \bibinfo{journal}{Nucl. Phys.} \textbf{\bibinfo{volume}{B399}},
  \bibinfo{pages}{654} (\bibinfo{year}{1993}{\natexlab{a}}),
  \eprint{hep-th/9203025}.

\bibitem[{\citenamefont{Ferrara et~al.}(1994)\citenamefont{Ferrara, Kounnas,
  and Zwirner}}]{Ferrara:1994kg}
\bibinfo{author}{\bibfnamefont{S.}~\bibnamefont{Ferrara}},
  \bibinfo{author}{\bibfnamefont{C.}~\bibnamefont{Kounnas}}, \bibnamefont{and}
  \bibinfo{author}{\bibfnamefont{F.}~\bibnamefont{Zwirner}},
  {``}\bibinfo{title}{{Mass formulae and natural hierarchy in string effective
  supergravities}},{''} \bibinfo{journal}{Nucl. Phys.}
  \textbf{\bibinfo{volume}{B429}}, \bibinfo{pages}{589} (\bibinfo{year}{1994}),
  \eprint{hep-th/9405188}.

\bibitem[{\citenamefont{Lopez et~al.}(1993{\natexlab{b}})\citenamefont{Lopez,
  Nanopoulos, and Zichichi}}]{Lopez:1993rm}
\bibinfo{author}{\bibfnamefont{J.~L.} \bibnamefont{Lopez}},
  \bibinfo{author}{\bibfnamefont{D.~V.} \bibnamefont{Nanopoulos}},
  \bibnamefont{and} \bibinfo{author}{\bibfnamefont{A.}~\bibnamefont{Zichichi}},
  {``}\bibinfo{title}{{Towards a unified string supergravity model}},{''}
  \bibinfo{journal}{Phys.Lett.} \textbf{\bibinfo{volume}{B319}},
  \bibinfo{pages}{451} (\bibinfo{year}{1993}{\natexlab{b}}),
  \eprint{hep-ph/9306226}.

\bibitem[{\citenamefont{Lopez et~al.}(1995{\natexlab{a}})\citenamefont{Lopez,
  Nanopoulos, and Zichichi}}]{Lopez:1994fz}
\bibinfo{author}{\bibfnamefont{J.~L.} \bibnamefont{Lopez}},
  \bibinfo{author}{\bibfnamefont{D.~V.} \bibnamefont{Nanopoulos}},
  \bibnamefont{and} \bibinfo{author}{\bibfnamefont{A.}~\bibnamefont{Zichichi}},
  {``}\bibinfo{title}{{Experimental consequences of one parameter no scale
  supergravity models}},{''} \bibinfo{journal}{Int.J.Mod.Phys.}
  \textbf{\bibinfo{volume}{A10}}, \bibinfo{pages}{4241}
  (\bibinfo{year}{1995}{\natexlab{a}}), \eprint{hep-ph/9408345}.

\bibitem[{\citenamefont{Lopez et~al.}(1995{\natexlab{b}})\citenamefont{Lopez,
  Nanopoulos, and Zichichi}}]{Lopez:1995hg}
\bibinfo{author}{\bibfnamefont{J.~L.} \bibnamefont{Lopez}},
  \bibinfo{author}{\bibfnamefont{D.~V.} \bibnamefont{Nanopoulos}},
  \bibnamefont{and} \bibinfo{author}{\bibfnamefont{A.}~\bibnamefont{Zichichi}},
  {``}\bibinfo{title}{{A String no scale supergravity model and its
  experimental consequences}},{''} \bibinfo{journal}{Phys.Rev.}
  \textbf{\bibinfo{volume}{D52}}, \bibinfo{pages}{4178}
  (\bibinfo{year}{1995}{\natexlab{b}}), \eprint{hep-ph/9502414}.

\bibitem[{\citenamefont{Lopez et~al.}(2007)\citenamefont{Lopez, Nanopoulos, and
  Zichichi}}]{superworld}
\bibinfo{author}{\bibfnamefont{J.~L.} \bibnamefont{Lopez}},
  \bibinfo{author}{\bibfnamefont{D.~V.} \bibnamefont{Nanopoulos}},
  \bibnamefont{and} \bibinfo{author}{\bibfnamefont{A.}~\bibnamefont{Zichichi}},
  \emph{\bibinfo{title}{Searching for the Superworld: A Volume in Honor of
  Antonino Zichichi on the Occasion of the Sixth Centenary Celebrations of the
  University of Turin, Italy}} (\bibinfo{publisher}{World Scientific Series in
  $20^{th}$ Century Physics}, \bibinfo{year}{2007}), vol. \bibinfo{volume}{39
  Part B.}, pp. \bibinfo{pages}{226--516}.

\bibitem[{\citenamefont{Witten}(1985)}]{Witten:1985xb}
\bibinfo{author}{\bibfnamefont{E.}~\bibnamefont{Witten}},
  {``}\bibinfo{title}{{Dimensional Reduction of Superstring Models}},{''}
  \bibinfo{journal}{Phys. Lett.} \textbf{\bibinfo{volume}{B155}},
  \bibinfo{pages}{151} (\bibinfo{year}{1985}).

\bibitem[{\citenamefont{Li et~al.}(1997)\citenamefont{Li, Lopez, and
  Nanopoulos}}]{Li:1997sk}
\bibinfo{author}{\bibfnamefont{T.-j.} \bibnamefont{Li}},
  \bibinfo{author}{\bibfnamefont{J.~L.} \bibnamefont{Lopez}}, \bibnamefont{and}
  \bibinfo{author}{\bibfnamefont{D.~V.} \bibnamefont{Nanopoulos}},
  {``}\bibinfo{title}{{Compactifications of M theory and their phenomenological
  consequences}},{''} \bibinfo{journal}{Phys.Rev.}
  \textbf{\bibinfo{volume}{D56}}, \bibinfo{pages}{2602} (\bibinfo{year}{1997}),
  \eprint{hep-ph/9704247}.

\bibitem[{\citenamefont{Beasley
  et~al.}(2009{\natexlab{a}})\citenamefont{Beasley, Heckman, and
  Vafa}}]{Beasley:2008dc}
\bibinfo{author}{\bibfnamefont{C.}~\bibnamefont{Beasley}},
  \bibinfo{author}{\bibfnamefont{J.~J.} \bibnamefont{Heckman}},
  \bibnamefont{and} \bibinfo{author}{\bibfnamefont{C.}~\bibnamefont{Vafa}},
  {``}\bibinfo{title}{{GUTs and Exceptional Branes in F-theory - I}},{''}
  \bibinfo{journal}{JHEP} \textbf{\bibinfo{volume}{01}}, \bibinfo{pages}{058}
  (\bibinfo{year}{2009}{\natexlab{a}}), \eprint{0802.3391}.

\bibitem[{\citenamefont{Beasley
  et~al.}(2009{\natexlab{b}})\citenamefont{Beasley, Heckman, and
  Vafa}}]{Beasley:2008kw}
\bibinfo{author}{\bibfnamefont{C.}~\bibnamefont{Beasley}},
  \bibinfo{author}{\bibfnamefont{J.~J.} \bibnamefont{Heckman}},
  \bibnamefont{and} \bibinfo{author}{\bibfnamefont{C.}~\bibnamefont{Vafa}},
  {``}\bibinfo{title}{{GUTs and Exceptional Branes in F-theory - II:
  Experimental Predictions}},{''} \bibinfo{journal}{JHEP}
  \textbf{\bibinfo{volume}{01}}, \bibinfo{pages}{059}
  (\bibinfo{year}{2009}{\natexlab{b}}), \eprint{0806.0102}.

\bibitem[{\citenamefont{Donagi and
  Wijnholt}(2008{\natexlab{a}})}]{Donagi:2008ca}
\bibinfo{author}{\bibfnamefont{R.}~\bibnamefont{Donagi}} \bibnamefont{and}
  \bibinfo{author}{\bibfnamefont{M.}~\bibnamefont{Wijnholt}},
  {``}\bibinfo{title}{{Model Building with F-Theory}},{''}
  (\bibinfo{year}{2008}{\natexlab{a}}), \eprint{0802.2969}.

\bibitem[{\citenamefont{Donagi and
  Wijnholt}(2008{\natexlab{b}})}]{Donagi:2008kj}
\bibinfo{author}{\bibfnamefont{R.}~\bibnamefont{Donagi}} \bibnamefont{and}
  \bibinfo{author}{\bibfnamefont{M.}~\bibnamefont{Wijnholt}},
  {``}\bibinfo{title}{{Breaking GUT Groups in F-Theory}},{''}
  (\bibinfo{year}{2008}{\natexlab{b}}), \eprint{0808.2223}.

\bibitem[{\citenamefont{Li et~al.}(2012{\natexlab{a}})\citenamefont{Li, Maxin,
  Nanopoulos, and Walker}}]{Li:2011xu}
\bibinfo{author}{\bibfnamefont{T.}~\bibnamefont{Li}},
  \bibinfo{author}{\bibfnamefont{J.~A.} \bibnamefont{Maxin}},
  \bibinfo{author}{\bibfnamefont{D.~V.} \bibnamefont{Nanopoulos}},
  \bibnamefont{and} \bibinfo{author}{\bibfnamefont{J.~W.}
  \bibnamefont{Walker}}, {``}\bibinfo{title}{{The Unification of Dynamical
  Determination and Bare Minimal Phenomenological Constraints in No-Scale
  \cal{F}- SU(5)}},{''} \bibinfo{journal}{Phys.Rev.}
  \textbf{\bibinfo{volume}{D85}}, \bibinfo{pages}{056007}
  (\bibinfo{year}{2012}{\natexlab{a}}), \eprint{1105.3988}.

\bibitem[{\citenamefont{Babu et~al.}(2008)\citenamefont{Babu, Gogoladze,
  Rehman, and Shafi}}]{Babu:2008ge}
\bibinfo{author}{\bibfnamefont{K.}~\bibnamefont{Babu}},
  \bibinfo{author}{\bibfnamefont{I.}~\bibnamefont{Gogoladze}},
  \bibinfo{author}{\bibfnamefont{M.~U.} \bibnamefont{Rehman}},
  \bibnamefont{and} \bibinfo{author}{\bibfnamefont{Q.}~\bibnamefont{Shafi}},
  {``}\bibinfo{title}{{Higgs Boson Mass, Sparticle Spectrum and Little
  Hierarchy Problem in Extended MSSM}},{''} \bibinfo{journal}{Phys.Rev.}
  \textbf{\bibinfo{volume}{D78}}, \bibinfo{pages}{055017}
  (\bibinfo{year}{2008}), \eprint{0807.3055}.

\bibitem[{\citenamefont{Martin}(2010)}]{Martin:2009bg}
\bibinfo{author}{\bibfnamefont{S.~P.} \bibnamefont{Martin}},
  {``}\bibinfo{title}{{Extra vector-like matter and the lightest Higgs scalar
  boson mass in low-energy supersymmetry}},{''} \bibinfo{journal}{Phys.Rev.}
  \textbf{\bibinfo{volume}{D81}}, \bibinfo{pages}{035004}
  (\bibinfo{year}{2010}), \eprint{0910.2732}.

\bibitem[{\citenamefont{Li et~al.}(2012{\natexlab{b}})\citenamefont{Li, Maxin,
  Nanopoulos, and Walker}}]{Li:2011ab}
\bibinfo{author}{\bibfnamefont{T.}~\bibnamefont{Li}},
  \bibinfo{author}{\bibfnamefont{J.~A.} \bibnamefont{Maxin}},
  \bibinfo{author}{\bibfnamefont{D.~V.} \bibnamefont{Nanopoulos}},
  \bibnamefont{and} \bibinfo{author}{\bibfnamefont{J.~W.}
  \bibnamefont{Walker}}, {``}\bibinfo{title}{{A Higgs Mass Shift to 125 GeV and
  A Multi-Jet Supersymmetry Signal: Miracle of the Flippons at the $\sqrt{s} =
  7$~TeV LHC}},{''} \bibinfo{journal}{Phys.Lett.}
  \textbf{\bibinfo{volume}{B710}}, \bibinfo{pages}{207}
  (\bibinfo{year}{2012}{\natexlab{b}}), \eprint{1112.3024}.

\bibitem[{\citenamefont{Li et~al.}(2012{\natexlab{c}})\citenamefont{Li, Maxin,
  Nanopoulos, and Walker}}]{Li:2012tr}
\bibinfo{author}{\bibfnamefont{T.}~\bibnamefont{Li}},
  \bibinfo{author}{\bibfnamefont{J.~A.} \bibnamefont{Maxin}},
  \bibinfo{author}{\bibfnamefont{D.~V.} \bibnamefont{Nanopoulos}},
  \bibnamefont{and} \bibinfo{author}{\bibfnamefont{J.~W.}
  \bibnamefont{Walker}}, {``}\bibinfo{title}{{Chanel $N^o5 ({\rm fb^{-1}})$:
  The Sweet Fragrance of SUSY}},{''} (\bibinfo{year}{2012}{\natexlab{c}}),
  \eprint{1205.3052}.

\bibitem[{\citenamefont{Li et~al.}(2011{\natexlab{f}})\citenamefont{Li, Maxin,
  Nanopoulos, and Walker}}]{Maxin:2011hy}
\bibinfo{author}{\bibfnamefont{T.}~\bibnamefont{Li}},
  \bibinfo{author}{\bibfnamefont{J.~A.} \bibnamefont{Maxin}},
  \bibinfo{author}{\bibfnamefont{D.~V.} \bibnamefont{Nanopoulos}},
  \bibnamefont{and} \bibinfo{author}{\bibfnamefont{J.~W.}
  \bibnamefont{Walker}}, {``}\bibinfo{title}{{The Ultrahigh jet multiplicity
  signal of stringy no-scale $\cal{F}$-$SU(5)$ at the $\sqrt{s}= 7$ TeV
  LHC}},{''} \bibinfo{journal}{Phys.Rev.} \textbf{\bibinfo{volume}{D84}},
  \bibinfo{pages}{076003} (\bibinfo{year}{2011}{\natexlab{f}}),
  \eprint{1103.4160}.

\bibitem[{\citenamefont{Li et~al.}(2011{\natexlab{g}})\citenamefont{Li, Maxin,
  Nanopoulos, and Walker}}]{Li:2011hr}
\bibinfo{author}{\bibfnamefont{T.}~\bibnamefont{Li}},
  \bibinfo{author}{\bibfnamefont{J.~A.} \bibnamefont{Maxin}},
  \bibinfo{author}{\bibfnamefont{D.~V.} \bibnamefont{Nanopoulos}},
  \bibnamefont{and} \bibinfo{author}{\bibfnamefont{J.~W.}
  \bibnamefont{Walker}}, {``}\bibinfo{title}{{Ultra High Jet Signals from
  Stringy No-Scale Supergravity}},{''} (\bibinfo{year}{2011}{\natexlab{g}}),
  \eprint{1103.2362}.

\bibitem[{\citenamefont{Li et~al.}(2012{\natexlab{d}})\citenamefont{Li, Maxin,
  Nanopoulos, and Walker}}]{Li:2011rp}
\bibinfo{author}{\bibfnamefont{T.}~\bibnamefont{Li}},
  \bibinfo{author}{\bibfnamefont{J.~A.} \bibnamefont{Maxin}},
  \bibinfo{author}{\bibfnamefont{D.~V.} \bibnamefont{Nanopoulos}},
  \bibnamefont{and} \bibinfo{author}{\bibfnamefont{J.~W.}
  \bibnamefont{Walker}}, {``}\bibinfo{title}{{Prospects for Discovery of
  Supersymmetric No-Scale F-SU(5) at The Once and Future LHC}},{''}
  \bibinfo{journal}{Nucl.Phys.} \textbf{\bibinfo{volume}{B859}},
  \bibinfo{pages}{96} (\bibinfo{year}{2012}{\natexlab{d}}), \eprint{1107.3825}.

\bibitem[{\citenamefont{Li et~al.}(2011{\natexlab{h}})\citenamefont{Li, Maxin,
  Nanopoulos, and Walker}}]{Li:2011fu}
\bibinfo{author}{\bibfnamefont{T.}~\bibnamefont{Li}},
  \bibinfo{author}{\bibfnamefont{J.~A.} \bibnamefont{Maxin}},
  \bibinfo{author}{\bibfnamefont{D.~V.} \bibnamefont{Nanopoulos}},
  \bibnamefont{and} \bibinfo{author}{\bibfnamefont{J.~W.}
  \bibnamefont{Walker}}, {``}\bibinfo{title}{{Has SUSY Gone Undetected in 9-jet
  Events? A Ten-Fold Enhancement in the LHC Signal Efficiency}},{''}
  (\bibinfo{year}{2011}{\natexlab{h}}), \eprint{1108.5169}.

\bibitem[{\citenamefont{Li et~al.}(2012{\natexlab{e}})\citenamefont{Li, Maxin,
  Nanopoulos, and Walker}}]{Li:2011xg}
\bibinfo{author}{\bibfnamefont{T.}~\bibnamefont{Li}},
  \bibinfo{author}{\bibfnamefont{J.~A.} \bibnamefont{Maxin}},
  \bibinfo{author}{\bibfnamefont{D.~V.} \bibnamefont{Nanopoulos}},
  \bibnamefont{and} \bibinfo{author}{\bibfnamefont{J.~W.}
  \bibnamefont{Walker}}, {``}\bibinfo{title}{{Natural Predictions for the Higgs
  Boson Mass and Supersymmetric Contributions to Rare Processes}},{''}
  \bibinfo{journal}{Phys.Lett.} \textbf{\bibinfo{volume}{B708}},
  \bibinfo{pages}{93} (\bibinfo{year}{2012}{\natexlab{e}}), \eprint{1109.2110}.

\bibitem[{\citenamefont{Li et~al.}(2011{\natexlab{i}})\citenamefont{Li, Maxin,
  Nanopoulos, and Walker}}]{Li:2011av}
\bibinfo{author}{\bibfnamefont{T.}~\bibnamefont{Li}},
  \bibinfo{author}{\bibfnamefont{J.~A.} \bibnamefont{Maxin}},
  \bibinfo{author}{\bibfnamefont{D.~V.} \bibnamefont{Nanopoulos}},
  \bibnamefont{and} \bibinfo{author}{\bibfnamefont{J.~W.}
  \bibnamefont{Walker}}, {``}\bibinfo{title}{{Profumo di SUSY: Suggestive
  Correlations in the ATLAS and CMS High Jet Multiplicity Data}},{''}
  (\bibinfo{year}{2011}{\natexlab{i}}), \eprint{1111.4204}.

\bibitem[{\citenamefont{Li et~al.}(2012{\natexlab{f}})\citenamefont{Li, Maxin,
  Nanopoulos, and Walker}}]{Li:2012hm}
\bibinfo{author}{\bibfnamefont{T.}~\bibnamefont{Li}},
  \bibinfo{author}{\bibfnamefont{J.~A.} \bibnamefont{Maxin}},
  \bibinfo{author}{\bibfnamefont{D.~V.} \bibnamefont{Nanopoulos}},
  \bibnamefont{and} \bibinfo{author}{\bibfnamefont{J.~W.}
  \bibnamefont{Walker}}, {``}\bibinfo{title}{{A Multi-Axis Best Fit to the
  Collider Supersymmetry Search: The Aroma of Stops and Gluinos at the
  $\sqrt{s}$ = 7 TeV LHC}},{''} (\bibinfo{year}{2012}{\natexlab{f}}),
  \eprint{1203.1918}.

\bibitem[{\citenamefont{Li et~al.}(2012{\natexlab{g}})\citenamefont{Li, Maxin,
  Nanopoulos, and Walker}}]{Li:2012ix}
\bibinfo{author}{\bibfnamefont{T.}~\bibnamefont{Li}},
  \bibinfo{author}{\bibfnamefont{J.~A.} \bibnamefont{Maxin}},
  \bibinfo{author}{\bibfnamefont{D.~V.} \bibnamefont{Nanopoulos}},
  \bibnamefont{and} \bibinfo{author}{\bibfnamefont{J.~W.}
  \bibnamefont{Walker}}, {``}\bibinfo{title}{{Non-trivial Supersymmetry
  Correlations between ATLAS and CMS Observations}},{''}
  (\bibinfo{year}{2012}{\natexlab{g}}), \eprint{1206.0293}.

\bibitem[{PAS(2011)}]{PAS-SUS-11-003}
{``}\bibinfo{title}{{Search for supersymmetry in all-hadronic events with
  $\alpha_{\rm T}$}},{''} (\bibinfo{year}{2011}), \bibinfo{note}{{CMS PAS
  SUS-11-003}}, \urlprefix\url{http://cdsweb.cern.ch}.

\bibitem[{\citenamefont{Aad et~al.}(2011{\natexlab{a}})}]{Aad:2011ib}
\bibinfo{author}{\bibfnamefont{G.}~\bibnamefont{Aad}} \bibnamefont{et~al.}
  (\bibinfo{collaboration}{ATLAS Collaboration}), {``}\bibinfo{title}{{Search
  for squarks and gluinos using final states with jets and missing transverse
  momentum with the ATLAS detector in $\sqrt{s}$ = 7 TeV proton-proton
  collisions}},{''} (\bibinfo{year}{2011}{\natexlab{a}}), \eprint{1109.6572}.

\bibitem[{\citenamefont{Aad et~al.}(2011{\natexlab{b}})}]{Aad:2011qa}
\bibinfo{author}{\bibfnamefont{G.}~\bibnamefont{Aad}} \bibnamefont{et~al.}
  (\bibinfo{collaboration}{Atlas Collaboration}), {``}\bibinfo{title}{{Search
  for new phenomena in final states with large jet multiplicities and missing
  transverse momentum using $\sqrt{s}$ = 7 TeV pp collisions with the ATLAS
  detector}},{''} \bibinfo{journal}{JHEP} \textbf{\bibinfo{volume}{1111}},
  \bibinfo{pages}{099} (\bibinfo{year}{2011}{\natexlab{b}}),
  \eprint{1110.2299}.

\bibitem[{\citenamefont{Aad et~al.}(2012{\natexlab{a}})}]{ATLAS-CONF-2012-037}
\bibinfo{author}{\bibfnamefont{G.}~\bibnamefont{Aad}} \bibnamefont{et~al.}
  (\bibinfo{collaboration}{ATLAS Collaboration}), {``}\bibinfo{title}{{Hunt for
  new phenomena using large jet multiplicities and missing transverse momentum
  with ATLAS in 4.7 ${\rm fb^{-1}}$ of $\sqrt{s} = 7$ TeV proton-proton
  collisions}},{''} \bibinfo{journal}{JHEP} \textbf{\bibinfo{volume}{1207}},
  \bibinfo{pages}{167} (\bibinfo{year}{2012}{\natexlab{a}}),
  \eprint{1206.1760}.

\bibitem[{ATL(2012{\natexlab{a}})}]{ATLAS-CONF-2012-033}
{``}\bibinfo{title}{{Search for squarks and gluinos with the ATLAS detector
  using final states with jets and missing transverse momentum and 4.7 ${\rm
  fb^{-1}}$ of $\sqrt{s}$ = 7 TeV proton-proton collision data}},{''}
  (\bibinfo{year}{2012}{\natexlab{a}}), \bibinfo{note}{{ATLAS-CONF-2012-033}},
  \eprint{1208.0949}, \urlprefix\url{http://cdsweb.cern.ch}.

\bibitem[{\citenamefont{Chatrchyan et~al.}(2012{\natexlab{a}})}]{:2012jx}
\bibinfo{author}{\bibfnamefont{S.}~\bibnamefont{Chatrchyan}}
  \bibnamefont{et~al.} (\bibinfo{collaboration}{CMS Collaboration}),
  {``}\bibinfo{title}{{Search for supersymmetry in hadronic final states using
  MT2 in $pp$ collisions at $\sqrt{s} = 7$ TeV}},{''}
  (\bibinfo{year}{2012}{\natexlab{a}}), \eprint{1207.1798}.

\bibitem[{\citenamefont{Li et~al.}(2012{\natexlab{h}})\citenamefont{Li, Maxin,
  Nanopoulos, and Walker}}]{Li:2012yd}
\bibinfo{author}{\bibfnamefont{T.}~\bibnamefont{Li}},
  \bibinfo{author}{\bibfnamefont{J.~A.} \bibnamefont{Maxin}},
  \bibinfo{author}{\bibfnamefont{D.~V.} \bibnamefont{Nanopoulos}},
  \bibnamefont{and} \bibinfo{author}{\bibfnamefont{J.~W.}
  \bibnamefont{Walker}}, {``}\bibinfo{title}{{Correlating LHCb $B_s^0 \to \mu^+
  \mu^-$ Results with the ATLAS-CMS Multijet Supersymmetry Search}},{''}
  \bibinfo{journal}{Europhysics.Lett.} \textbf{\bibinfo{volume}{In Press}}
  (\bibinfo{year}{2012}{\natexlab{h}}), \eprint{1206.2633}.

\bibitem[{\citenamefont{Li et~al.}(2012{\natexlab{i}})\citenamefont{Li, Maxin,
  Nanopoulos, and Walker}}]{Li:2012jf}
\bibinfo{author}{\bibfnamefont{T.}~\bibnamefont{Li}},
  \bibinfo{author}{\bibfnamefont{J.~A.} \bibnamefont{Maxin}},
  \bibinfo{author}{\bibfnamefont{D.~V.} \bibnamefont{Nanopoulos}},
  \bibnamefont{and} \bibinfo{author}{\bibfnamefont{J.~W.}
  \bibnamefont{Walker}}, {``}\bibinfo{title}{{A 125.5 GeV Higgs Boson in ${\cal
  F}$-$SU(5)$: Imminently Observable Proton Decay, A 130 GeV Gamma-ray Line,
  and SUSY Multijets \& Light Stops at the LHC8}},{''}
  (\bibinfo{year}{2012}{\natexlab{i}}), \eprint{1208.1999}.

\bibitem[{\citenamefont{Li et~al.}(2012{\natexlab{j}})\citenamefont{Li, Maxin,
  Nanopoulos, and Walker}}]{Li:2012qv}
\bibinfo{author}{\bibfnamefont{T.}~\bibnamefont{Li}},
  \bibinfo{author}{\bibfnamefont{J.~A.} \bibnamefont{Maxin}},
  \bibinfo{author}{\bibfnamefont{D.~V.} \bibnamefont{Nanopoulos}},
  \bibnamefont{and} \bibinfo{author}{\bibfnamefont{J.~W.}
  \bibnamefont{Walker}}, {``}\bibinfo{title}{{Testing No-Scale ${\cal
  F}$-$SU(5)$: A 125 GeV Higgs Boson and SUSY at the 8 TeV LHC}},{''}
  \bibinfo{journal}{Phys.Lett.} \textbf{\bibinfo{volume}{B718}},
  \bibinfo{pages}{70} (\bibinfo{year}{2012}{\natexlab{j}}), \eprint{1207.1051}.

\bibitem[{\citenamefont{Li et~al.}(2012{\natexlab{k}})\citenamefont{Li, Maxin,
  Nanopoulos, and Walker}}]{Li:2012mr}
\bibinfo{author}{\bibfnamefont{T.}~\bibnamefont{Li}},
  \bibinfo{author}{\bibfnamefont{J.~A.} \bibnamefont{Maxin}},
  \bibinfo{author}{\bibfnamefont{D.~V.} \bibnamefont{Nanopoulos}},
  \bibnamefont{and} \bibinfo{author}{\bibfnamefont{J.~W.}
  \bibnamefont{Walker}}, {``}\bibinfo{title}{{Primordial Synthesis: F-SU(5)
  SUSY Multijets, 145-150 GeV LSP, Proton \& Rare Decays, 125 GeV Higgs Boson,
  and WMAP7}},{''} (\bibinfo{year}{2012}{\natexlab{k}}), \eprint{1210.3011}.

\bibitem[{\citenamefont{Artuso et~al.}(2009)\citenamefont{Artuso, Barberio, and
  Stone}}]{Artuso:2009jw}
\bibinfo{author}{\bibfnamefont{M.}~\bibnamefont{Artuso}},
  \bibinfo{author}{\bibfnamefont{E.}~\bibnamefont{Barberio}}, \bibnamefont{and}
  \bibinfo{author}{\bibfnamefont{S.}~\bibnamefont{Stone}},
  {``}\bibinfo{title}{{$B$ Meson Decays}},{''} \bibinfo{journal}{PMC Phys.}
  \textbf{\bibinfo{volume}{A3}}, \bibinfo{pages}{3} (\bibinfo{year}{2009}),
  \eprint{0902.3743}.

\bibitem[{\citenamefont{Misiak}(2011)}]{Misiak:2010dz}
\bibinfo{author}{\bibfnamefont{M.}~\bibnamefont{Misiak}},
  {``}\bibinfo{title}{{QCD challenges in radiative B decays}},{''}
  \bibinfo{journal}{AIP Conf.Proc.} \textbf{\bibinfo{volume}{1317}},
  \bibinfo{pages}{276} (\bibinfo{year}{2011}), \eprint{1010.4896}.

\bibitem[{\citenamefont{Becher and Neubert}(2007)}]{Becher:2006pu}
\bibinfo{author}{\bibfnamefont{T.}~\bibnamefont{Becher}} \bibnamefont{and}
  \bibinfo{author}{\bibfnamefont{M.}~\bibnamefont{Neubert}},
  {``}\bibinfo{title}{{Analysis of Br$(\overline{B} \to X_s \gamma)$ at NNLO
  with a cut on photon energy}},{''} \bibinfo{journal}{Phys. Rev. Lett.}
  \textbf{\bibinfo{volume}{98}}, \bibinfo{pages}{022003}
  (\bibinfo{year}{2007}), \eprint{hep-ph/0610067}.

\bibitem[{\citenamefont{Aaij et~al.}(2012)}]{Aaij:2012ac}
\bibinfo{author}{\bibfnamefont{R.}~\bibnamefont{Aaij}} \bibnamefont{et~al.}
  (\bibinfo{collaboration}{LHCb collaboration}), {``}\bibinfo{title}{{Strong
  constraints on the rare decays $B_s \to \mu^+ \mu^-$ and $B^0 \to \mu^+
  \mu^-$}},{''} \bibinfo{journal}{Phys.Rev.Lett.}
  \textbf{\bibinfo{volume}{108}}, \bibinfo{pages}{231801}
  (\bibinfo{year}{2012}), \eprint{1203.4493}.

\bibitem[{\citenamefont{Bandurin}(2011)}]{Bandurin:2011sh}
\bibinfo{author}{\bibfnamefont{D.}~\bibnamefont{Bandurin}}
  (\bibinfo{collaboration}{D0 and CDF Collaborations}),
  {``}\bibinfo{title}{{QCD measurements at the Tevatron}},{''}
  (\bibinfo{year}{2011}), \eprint{1112.0051}.

\bibitem[{\citenamefont{Belanger et~al.}(2011)\citenamefont{Belanger, Boudjema,
  Brun, Pukhov, Rosier-Lees et~al.}}]{Belanger:2010gh}
\bibinfo{author}{\bibfnamefont{G.}~\bibnamefont{Belanger}},
  \bibinfo{author}{\bibfnamefont{F.}~\bibnamefont{Boudjema}},
  \bibinfo{author}{\bibfnamefont{P.}~\bibnamefont{Brun}},
  \bibinfo{author}{\bibfnamefont{A.}~\bibnamefont{Pukhov}},
  \bibinfo{author}{\bibfnamefont{S.}~\bibnamefont{Rosier-Lees}},
  \bibnamefont{et~al.}, {``}\bibinfo{title}{{Indirect search for dark matter
  with micrOMEGAs2.4}},{''} \bibinfo{journal}{Comput.Phys.Commun.}
  \textbf{\bibinfo{volume}{182}}, \bibinfo{pages}{842} (\bibinfo{year}{2011}),
  \eprint{1004.1092}.

\bibitem[{\citenamefont{Djouadi et~al.}(2007)\citenamefont{Djouadi, Kneur, and
  Moultaka}}]{Djouadi:2002ze}
\bibinfo{author}{\bibfnamefont{A.}~\bibnamefont{Djouadi}},
  \bibinfo{author}{\bibfnamefont{J.-L.} \bibnamefont{Kneur}}, \bibnamefont{and}
  \bibinfo{author}{\bibfnamefont{G.}~\bibnamefont{Moultaka}},
  {``}\bibinfo{title}{{SuSpect: A Fortran code for the supersymmetric and Higgs
  particle spectrum in the MSSM}},{''} \bibinfo{journal}{Comput. Phys. Commun.}
  \textbf{\bibinfo{volume}{176}}, \bibinfo{pages}{426} (\bibinfo{year}{2007}),
  \eprint{hep-ph/0211331}.

\bibitem[{\citenamefont{Hewett et~al.}(2012)\citenamefont{Hewett, Weerts,
  Brock, Butler, Casey et~al.}}]{Hewett:2012ns}
\bibinfo{author}{\bibfnamefont{J.}~\bibnamefont{Hewett}},
  \bibinfo{author}{\bibfnamefont{H.}~\bibnamefont{Weerts}},
  \bibinfo{author}{\bibfnamefont{R.}~\bibnamefont{Brock}},
  \bibinfo{author}{\bibfnamefont{J.}~\bibnamefont{Butler}},
  \bibinfo{author}{\bibfnamefont{B.}~\bibnamefont{Casey}},
  \bibnamefont{et~al.}, {``}\bibinfo{title}{{Fundamental Physics at the
  Intensity Frontier}},{''} (\bibinfo{year}{2012}), \eprint{1205.2671}.

\bibitem[{\citenamefont{Aprile et~al.}(2012)}]{Aprile:2012nq}
\bibinfo{author}{\bibfnamefont{E.}~\bibnamefont{Aprile}} \bibnamefont{et~al.}
  (\bibinfo{collaboration}{XENON100 Collaboration}), {``}\bibinfo{title}{{Dark
  Matter Results from 225 Live Days of XENON100 Data}},{''}
  (\bibinfo{year}{2012}), \eprint{1207.5988}.

\bibitem[{\citenamefont{Li et~al.}(2011{\natexlab{j}})\citenamefont{Li, Maxin,
  Nanopoulos, and Walker}}]{Li:2011in}
\bibinfo{author}{\bibfnamefont{T.}~\bibnamefont{Li}},
  \bibinfo{author}{\bibfnamefont{J.~A.} \bibnamefont{Maxin}},
  \bibinfo{author}{\bibfnamefont{D.~V.} \bibnamefont{Nanopoulos}},
  \bibnamefont{and} \bibinfo{author}{\bibfnamefont{J.~W.}
  \bibnamefont{Walker}}, {``}\bibinfo{title}{{The Race for Supersymmetric Dark
  Matter at XENON100 and the LHC: Stringy Correlations from No-Scale
  \cal{F}-SU(5)}},{''} (\bibinfo{year}{2011}{\natexlab{j}}),
  \eprint{1106.1165}.

\bibitem[{\citenamefont{Weniger}(2012)}]{Weniger:2012tx}
\bibinfo{author}{\bibfnamefont{C.}~\bibnamefont{Weniger}},
  {``}\bibinfo{title}{{A Tentative Gamma-Ray Line from Dark Matter Annihilation
  at the Fermi Large Area Telescope}},{''} (\bibinfo{year}{2012}),
  \eprint{1204.2797}.

\bibitem[{\citenamefont{Bringmann et~al.}(2012)\citenamefont{Bringmann, Huang,
  Ibarra, Vogl, and Weniger}}]{Bringmann:2012vr}
\bibinfo{author}{\bibfnamefont{T.}~\bibnamefont{Bringmann}},
  \bibinfo{author}{\bibfnamefont{X.}~\bibnamefont{Huang}},
  \bibinfo{author}{\bibfnamefont{A.}~\bibnamefont{Ibarra}},
  \bibinfo{author}{\bibfnamefont{S.}~\bibnamefont{Vogl}}, \bibnamefont{and}
  \bibinfo{author}{\bibfnamefont{C.}~\bibnamefont{Weniger}},
  {``}\bibinfo{title}{{Fermi LAT Search for Internal Bremsstrahlung Signatures
  from Dark Matter Annihilation}},{''} (\bibinfo{year}{2012}),
  \eprint{1203.1312}.

\bibitem[{\citenamefont{Ackermann et~al.}(2012)}]{Ackermann:2012qk}
\bibinfo{author}{\bibfnamefont{M.}~\bibnamefont{Ackermann}}
  \bibnamefont{et~al.} (\bibinfo{collaboration}{LAT Collaboration}),
  {``}\bibinfo{title}{{Fermi LAT Search for Dark Matter in Gamma-ray Lines and
  the Inclusive Photon Spectrum}},{''} \bibinfo{journal}{Phys.Rev.}
  \textbf{\bibinfo{volume}{D86}}, \bibinfo{pages}{022002}
  (\bibinfo{year}{2012}), \eprint{1205.2739}.

\bibitem[{\citenamefont{Hektor et~al.}(2012)\citenamefont{Hektor, Raidal, and
  Tempel}}]{Hektor:2012kc}
\bibinfo{author}{\bibfnamefont{A.}~\bibnamefont{Hektor}},
  \bibinfo{author}{\bibfnamefont{M.}~\bibnamefont{Raidal}}, \bibnamefont{and}
  \bibinfo{author}{\bibfnamefont{E.}~\bibnamefont{Tempel}},
  {``}\bibinfo{title}{{An evidence for indirect detection of dark matter from
  galaxy clusters in Fermi-LAT data}},{''} (\bibinfo{year}{2012}),
  \eprint{1207.4466}.

\bibitem[{\citenamefont{Group}(2012)}]{LEPEWWG}
\bibinfo{author}{\bibfnamefont{L.~E.~W.} \bibnamefont{Group}}
  (\bibinfo{year}{2012}), \bibinfo{note}{{Blue Band Higgs $\chi^2$ plot}},
  \urlprefix\url{http://lepewwg.web.cern.ch/LEPEWWG/}.

\bibitem[{\citenamefont{Malaescu}(2012)}]{Malaescu:2012iq}
\bibinfo{author}{\bibfnamefont{B.}~\bibnamefont{Malaescu}},
  {``}\bibinfo{title}{{Evaluation of $\alpha_s$ using the ATLAS inclusive jet
  cross-section data}},{''} (\bibinfo{year}{2012}), \eprint{1210.1383}.

\bibitem[{\citenamefont{Chatrchyan et~al.}(2012{\natexlab{b}})}]{:2012cz}
\bibinfo{author}{\bibfnamefont{S.}~\bibnamefont{Chatrchyan}}
  \bibnamefont{et~al.} (\bibinfo{collaboration}{The CMS Collaboration}),
  {``}\bibinfo{title}{{Measurement of the top-quark mass in $t \bar{t}$ events
  with lepton+jets final states in pp collisions at $\sqrt{s} = 7$ TeV}},{''}
  (\bibinfo{year}{2012}{\natexlab{b}}), \eprint{1209.2319}.

\bibitem[{\citenamefont{Aad et~al.}(2012{\natexlab{b}})}]{ATLAS:2012aj}
\bibinfo{author}{\bibfnamefont{G.}~\bibnamefont{Aad}} \bibnamefont{et~al.}
  (\bibinfo{collaboration}{ATLAS Collaboration}),
  {``}\bibinfo{title}{{Measurement of the top quark mass with the template
  method in the $t \bar{t} \to$ lepton + jets channel using ATLAS data}},{''}
  \bibinfo{journal}{Eur.Phys.J.} \textbf{\bibinfo{volume}{C72}},
  \bibinfo{pages}{2046} (\bibinfo{year}{2012}{\natexlab{b}}),
  \eprint{1203.5755}.

\bibitem[{ATL(2012{\natexlab{b}})}]{ATLAS-CONF-2012-030}
{``}\bibinfo{title}{{Determination of the Top Quark Mass with a Template Method
  in the All-Hadronic Decay Channel using 2.04 ${\rm fb^{-1}}$ of ATLAS
  Data}},{''} (\bibinfo{year}{2012}{\natexlab{b}}),
  \bibinfo{note}{{ATLAS-CONF-2012-030}}, \urlprefix\url{http://cdsweb.cern.ch}.

\bibitem[{ATL(2012{\natexlab{c}})}]{ATLAS-CONF-2012-082}
{``}\bibinfo{title}{{Top-quark mass measurement in the $e \mu$ channel using
  the $m_{T2}$ variable at ATLAS}},{''} (\bibinfo{year}{2012}{\natexlab{c}}),
  \bibinfo{note}{{ATLAS-CONF-2012-082}}, \urlprefix\url{http://cdsweb.cern.ch}.

\bibitem[{\citenamefont{Ellis et~al.}(1996)\citenamefont{Ellis, Lopez, and
  Nanopoulos}}]{Ellis:1995at}
\bibinfo{author}{\bibfnamefont{J.~R.} \bibnamefont{Ellis}},
  \bibinfo{author}{\bibfnamefont{J.~L.} \bibnamefont{Lopez}}, \bibnamefont{and}
  \bibinfo{author}{\bibfnamefont{D.~V.} \bibnamefont{Nanopoulos}},
  {``}\bibinfo{title}{{Lowering $\alpha_s$ by flipping SU(5)}},{''}
  \bibinfo{journal}{Phys.Lett.} \textbf{\bibinfo{volume}{B371}},
  \bibinfo{pages}{65} (\bibinfo{year}{1996}), \eprint{hep-ph/9510246}.

\end{thebibliography}

\end{document}